\documentclass[aps,10pt,prd,groupedaddress,nofootinbib,notitlepage,eqsecnum,preprintnumbers]{revtex4-1}

\usepackage[utf8]{inputenc}
\usepackage{hyperref}
\usepackage{xcolor}
\usepackage{graphicx}
\usepackage{amsmath,amssymb}
\usepackage{bm}
\usepackage{floatrow}
\usepackage{capt-of}
\usepackage[font=small,labelfont=bf,tableposition=top]{caption}
\usepackage{booktabs}
\usepackage{varwidth}
\usepackage[export]{adjustbox}
\newfloatcommand{capbtabbox}{table}[][\FBwidth]

\DeclareCaptionLabelFormat{andtable}{\tablename~\thetable \&  \tablename~\thetable}

\begin{document}

\title{Planck residuals anomaly as a fingerprint of alternative scenarios to inflation}

\author{\textsc{Guillem Dom\`enech$^{a}$}}
		\email{{domenech}@{thphys.uni-heidelberg.de}}
\author{\textsc{Xingang Chen$^{b}$}}
\author{\textsc{Marc Kamionkowski$^{c}$} }
\author{\textsc{Abraham Loeb$^{b}$} }

\affiliation{$^{a}$\small{Institut f\"ur Theoretische Physik, Ruprecht-Karls-Universit\"at Heidelberg, Philosophenweg 16, 69120 Heidelberg, Germany}\\
          $^{b}$\small{Harvard-Smithsonian Center for Astrophysics,
			60 Garden Street, Cambridge, MA 02138, USA}\\
			$^{c}$\small{Department of Physics \& Astronomy, Johns Hopkins
			University, 3400 N.\ Charles St., Baltimore, MD 21218, USA}
		}

\begin{abstract}
Planck's residuals of the CMB temperature power spectrum present a curious oscillatory shape that resembles an extra smoothing effect of lensing and is the source of the lensing anomaly. The smoothing effect of lensing to the CMB temperature power spectrum is, to some extent, degenerate with oscillatory modulations of the primordial power spectrum, in particular if the frequency is close to that of the acoustic peaks. We consider the possibility that the lensing anomaly reported by the latest Planck 2018 results may be hinting at an oscillatory modulation generated by a massive scalar field during an alternative scenario to inflation or by a sharp feature during inflation. We use the full TTTEEE+low E CMB likelihood from Planck to derive constraints on these two types of models. We obtain that in both cases the $A_L$ anomaly is mildly reduced to slightly less than $2\sigma$, to be compared with the $2.8\sigma$ deviation from $A_L=1$ in $\Lambda$CDM. Although the oscillatory features are not able to satisfactorily ease the lensing anomaly, we find that the oscillatory modulation generated during an alternative scenario alone, i.e. with $A_L=1$, presents the lowest value of $\chi^2$, with $\Delta\chi^2=-13$ compared to $\Lambda$CDM. Furthermore, the Akaike Information Criterion suggests that such an oscillation constitutes an attractive candidate since it has a value $\Delta{\rm AIC}=-5$ with respect to $\Lambda$CDM, comparable to the $A_L$ parameter. We also obtain that the equation of state parameter in the alternative scenario is given at $1\sigma$ by
$w=0.13\pm0.17$. Interestingly, the matter bounce and radiation bounce scenarios are compatible with our results. We discuss how these models of oscillatory features can be tested with future observations.
\end{abstract}


 \maketitle

\section{Introduction}

Observations of the cosmic microwave background (CMB) showed that the primordial spectrum of fluctuations is near scale invariant, adiabatic and gaussian \cite{Aghanim:2018eyx}. The leading paradigm to explain these facts is inflation \cite{Brout:1977ix,Starobinsky:1980te,Guth:1980zm,Sato:1980yn,Linde:1981mu,Albrecht:1982wi,
Mukhanov:1981xt,Mukhanov:1990me}, a period of accelerated expansion in the primeval universe. However, one should not ignore other distant competitors to the throne. Scenarios such as the ekpyrotic universes \cite{Khoury:2001wf,Ijjas:2018qbo,Brandenberger:2020eyf}, matter contraction/bounce \cite{Wands:1998yp}, radiation contraction/bounce \cite{Boyle:2018tzc},  string gas cosmology \cite{Brandenberger:1988aj} and pre-big-bang cosmlogy \cite{Gasperini:1992em} may be able to generate a primordial spectrum compatible with observations, although they face more theoretical problems (see Ref.~\cite{Battefeld:2014uga,Brandenberger:2016vhg} for general reviews). Even if these candidates are currently not on equal footing, it would be desirable to have a robust observable able to tell them apart.

A promising candidate to discriminate between models of the primordial universe is the imprint in the primordial density fluctuations left by scalar massive fields \cite{Chen:2011zf,Chen:2011tu,Chen:2015lza,Chen:2018cgg}. Already present in the standard model of particle physics with the Higgs, scalar fields are also ubiquitous in higher dimensional theories, like braneworld cosmology and string theory, after dimensional reduction \cite{Fujii:2003pa}. In particular, such dimensional compactification quite generically leads to many heavy fields with constant masses in the resulting effective theory \cite{Baumann:2014nda}. This means that, in an expanding or contracting universe we expect the presence of many fields with masses larger than the scale of the horizon. We will study scalar fields in this paper.

The basic idea of the signature is as follows. The mass $M$ of a heavy scalar field is directly related to the frequency of the classical oscillations of the field, if somehow excited above the minimum of the potential, as well as related to the frequency of the quantum fluctuations in a mass dominated regime if this field is not excited classically. If the field is excited with a displacement from the minimum or with a non-trivial state for the mode functions, the frequency will be imprinted in the primordial spectrum,
assuming a direct coupling to the field responsible for generating the curvature perturbations. In the end, the modifications to the curvature perturbations contribute most when a given mode with physical wavenumber $k/a$ crosses the scale corresponding to $M$, where $a$ is the scale factor of the universe. In other words, the oscillations of the massive scalar field have to be evaluated at $k=Ma$. If the mass $M$ is constant, the resulting oscillatory features in the primordial fluctuations directly probe the evolution of the primordial universe through the scale factor $a$ \cite{Chen:2011zf,Chen:2011tu,Chen:2015lza,Chen:2018cgg}.
These massive fields are referred to as primordial standard clocks (PSCs).

To be more concrete, the oscillatory modulations acquire a $k$-dependent frequency and are roughly given by $\sin(\omega k^\gamma)$, where $\omega$ and $\gamma$ are constant. The exponent $\gamma$ is directly related to the time dependence of the scale factor. On one hand, for an inflationary phase we have that $\gamma\ll1$, reaching a logarithmic running\footnote{Note that a similar running in $k$ is obtained by trans-planckian modulations due to a new physics hypersurface ($\gamma\ll1$) \cite{Easther:2005yr,Brandenberger:2012aj} and in the boundary effective field theory ($\gamma=1$) \cite{Easther:2005yr}. However, the frequency of the oscillatory feature due to trans-planckian modulations is proportional to the initial conditions. This results in a frequency higher than needed to explain the lensing anomaly \cite{Domenech:2019cyh}.} in $k$ for exact de Sitter expansion \cite{Chen:2011zf}. On the other hand, for a contracting phase (pre-bounce) one finds $\gamma\gtrsim 1$ \cite{Chen:2011zf,Chen:2018cgg}. The exact constant frequency case, i.e. $\gamma=1$, is typical of sharp transitions \cite{Chen:2010xka,Chluba:2015bqa}. Thus, oscillatory features in the primordial spectrum, especially the $k$-dependence in the frequency, might be a potential discriminator for models of the very early universe.
However, although there are reasons to believe the existence of many heavy fields with constant masses, it is also interesting to consider more general cases and examine the effects generated by time-dependent clock frequencies, namely non-standard clocks \cite{Chen:2011zf,Huang:2016quc,Domenech:2018bnf}. To what extent these can cause confusion among different scenarios and how they can be further distinguished are interesting questions for future studies.

Recently, the latest study of CMB data by Planck pointed out that there seems to be $10\%$ more lensing of the temperature power spectrum than expected in standard $\Lambda$CDM \cite{Aghanim:2018eyx}. This is due to the fact that Planck's residuals of the CMB temperature power spectrum present an interesting oscillatory shape that resembles an extra smoothing effect of lensing. On one hand, this roughly $2.8\sigma$ tension with $\Lambda$CDM is mainly driven by the TT power spectrum when including the smearing of the acoustic peaks \cite{Aghanim:2018eyx,Simard:2017xtw,Motloch:2019gux} (TE and EE respectively prefer slighlty lower and higher lensing). On the other hand, studies of the CMB lensing power spectrum, without considering the smoothing of the acoustic peaks, are in good agreement with $\Lambda$CDM expectations \cite{Simard:2017xtw,Motloch:2019gux,Aghanim:2018oex}. Furthermore, the SPTpol data reveals a preference for less smoothing of the acoustic peaks although consistent with $\Lambda$CDM within $1.4\sigma$ \cite{Simard:2017xtw,Henning:2017nuy,Bianchini:2019vxp,Chudaykin:2020acu}. Also, it has been shown that reionization blurring cannot acount for the lensing anomaly \cite{Fidler:2019mny}. Thus, if the anomaly remains, it could be due to new physics which mimic the lensing smoothing of the acoustic peaks in the power spectrum. For example, it could be due to cold dark matter isocurvature perturbations \cite{Munoz:2015fdv,Smith:2017ndr,Aghanim:2018eyx}, which are tightly constrained by their effects to the trispectrum \cite{Smith:2017ndr}, or an oscillatory feature in the primordial spectrum with a frequency comparable to the acoustic peaks \cite{Aghanim:2018eyx,Domenech:2019cyh}. Regarding the latter, it should be noted that Refs.~\cite{Hazra:2014jwa,Hazra:2018opk} already pointed out that oscillations in the primordial power spectrum are degenerate with the smoothing effects of lensing. In any case, even if the oscillations may not provide a complete explanation to the anomaly, such seemingly extra lensing provides a window to find a potential candidate of oscillatory features in the primordial spectrum.

Analysis by the Planck team found no evidence of an oscillatory feature with constant frequency and amplitude that eases the so-called $A_L$ anomaly \cite{Calabrese:2008rt,Aghanim:2018eyx,Akrami:2018odb}. Only when the oscillatory feature has a fine tuned k-dependent envelope \cite{Akrami:2018odb} the tension is removed. However, as was argued in Ref.~\cite{Domenech:2019cyh}, general models of inflation and its alternatives do not predict a constant amplitude nor frequency. Furthermore, Ref.~\cite{Domenech:2019cyh} provided a potential candidate with a constant frequency but $k$ dependent envelope that may reduce the tension. It should be noted that Ref.~\cite{Domenech:2019cyh} did not consider $k$ dependent frequencies as they are unable to fit the acoustic peaks in the whole range of CMB multipoles. This was clearly an oversimplification since the anomaly is more pronounced for $1100\lesssim\ell \lesssim  2000$ and a fit of a few peaks may be enough to reduce the tension. In this paper, we will push forward this claim by fitting the latest CMB 2018 data by Planck \cite{Aghanim:2018eyx} with the inflationary model of Ref.~\cite{Domenech:2019cyh}. Furthermore, we will use the chance to test whether oscillatory features generated in contracting scenarios from Ref.~\cite{Chen:2018cgg} yield a better fit to the data. Thus, we will provide the first comparison of inflation with one of its alternatives through a direct measurement of the scale factor evolution.

The study of oscillatory features in the primordial spectrum is particularly relevant for early universe physics considering future observational prospects (see e.g.~Ref.~\cite{Chen:2010xka,Chluba:2015bqa} for extensive reviews on feature models in inflation models). The constraints on such features are expected to improve by polarization data of future CMB experiments and by large scale structure surveys \cite{Huang:2012mr,Meerburg:2015owa,Chen:2016vvw,Ballardini:2016hpi,Chen:2016zuu,Xu:2016kwz,Ballardini:2017qwq}. Moreover, it has been recently argued that oscillatory features in the primordial power spectrum can be modelled to non-linear evolution of matter and could be seen in, e.g., the baryon acoustic oscillations (BAO) \cite{Vlah:2015zda,Beutler:2019ojk,Vasudevan:2019ewf,Ballardini:2019tuc}. Thus, large-scale structure observations might largely improve CMB constraints on oscillatory features.

The paper is organized as follows. In Sec.~\ref{sec:imprint} we briefly review the mechanisms to generate oscillatory features in the primordial power spectrum in general scenarios. We also present why they may ease the $A_L$ anomaly. In Sec.~\ref{sec:results}, we contrast the templates against the latest CMB temperature and polarization data. We summarize our work and discussion in Sec.~\ref{sec:conclusions}. In App.~\ref{app:T1extra} we provide complementary results.

\section{Imprint of oscillating massive fields in the primordial fluctuations \label{sec:imprint}}

We start by deriving the main predictions of PSCs as fingerprints of the evolution of the primordial universe. We will follow a rather heuristic derivation and we refer the reader to Refs.~\cite{Chen:2011zf,Chen:2011tu,Chen:2015lza,Chen:2018cgg} for details. First, let us consider that the geometry is described by a flat FLRW metric, namely
\begin{align}
ds^2=a(\tau)^2\left(-d\tau^2+\delta_{ij}dx^idx^j\right)\,,
\end{align}
and that the universe follows a power-law expansion given by
\begin{align}\label{eq:powerlaw}
a(\tau)=a_0\left(\frac{\tau}{\tau_0}\right)^\alpha
=a_0 \left( \frac{t}{t_0} \right)^p
\quad
{\rm and}\quad {\cal H}\equiv\frac{a'}{a}=\frac{\alpha}{\tau}\,,
\end{align}
where ${\cal H}$ is the conformal Hubble parameter and $dt=a d\tau$.
The power-law index $\alpha$ is related to the equation of state of matter $w=P/\rho$, where $P$ stands for pressure and $\rho$ for energy density, by
\begin{align}
\alpha=\frac{2}{(1 + 3 w)}\,.
\end{align}
Later in the paper, we will also introduce a parameter $\gamma$, equivalent to $\alpha$ or $p$, in the data analysis template. The relations between these parameters are
\begin{align}
\gamma = 1 + \frac{1}{\alpha} = \frac{1}{p} ~.
\end{align}
On one hand, for an expanding background we have $p<0$ or $p>1$ (or, equivalently, $w<-1/3$, $\alpha<0$, $\gamma<1$).
In particular, for inflation,  $|p|\gg 1$ ($w\sim-1$, $\alpha\sim-1$, $|\gamma |\ll 1$);
for slowly expanding scenarios, $p<0$ and $|p|\ll 1$ ($|w|\gg 1$, $|\alpha|\ll 1$, $\gamma\ll -1$).
On the other hand, for contracting backgrounds, we have $0<p<1$  (or, equivalently, $w>-1/3$, $\alpha>0$, $\gamma>1$).
In particular, for the ekpyrotic universe, $0<p\ll1$ ($w\gg 1$, $0<\alpha\ll1$, $\gamma\gg 1$); for matter contraction/bounce scenario, $p=2/3$ ($w=0$, $\alpha=2$, $\gamma=3/2$); for radiation contraction/bounce scenario, $p=1/2$ ($w=1/3$, $\alpha=1$, $\gamma=2$).
In any of the cases at hand, conformal time always runs from $-\infty<\tau<0$ but $t$ runs in different domains depending on scenarios.

We also consider, for simplicity, that the matter content is composed of two scalar fields. One field $\varphi$ dominates the energy density of the universe and yields Eq.~\eqref{eq:powerlaw} and is responsible for generating the curvature perturbation ${\cal R}$. The other field $\sigma$ is a massive spectator field with mass larger than the mass scale of the horizon. Note that, for non-inflation models, the mass scale of the horizon is generically not the same as the Hubble parameter and it is also highly time-dependent \cite{Chen:2011zf}. Consequently, for example, a field initially considered to be heavy can later become relatively light due to the increase in the horizon mass scale.

The Lagrangian of the model is given by
\begin{align}
{\cal L}_\sigma=-\frac{1}{2}g^{\mu\nu}\partial_\mu\sigma\partial_\nu\sigma-\frac{1}{2}M^2\sigma^2 + {\rm interactions}
\,.
\end{align}
Using the uniform-$\varphi$ gauge, in the perturbation theory \cite{Chen:2011zf,Noumi:2012vr,Tong:2017iat,Chen:2018cgg}, the scalar field $\sigma$ and its fluctuations $\delta\sigma$ can directly couple to the curvature perturbation, e.g. by some interaction terms such as
\begin{align}
{\cal L}^{(2)}_{\rm int,\sigma}=\lambda_1 \sigma {\cal R}^2\quad{\rm and}\quad
{\delta\cal L}^{(2)}_{\rm int,\delta\sigma}=\lambda_2 \delta\sigma {\cal R}\,,
\end{align}
where $\lambda_1$ and $\lambda_2$ parametrize the strength of two interactions, responsible for classical and quantum PSC cases, respectively. We will treat the massive field $\delta\sigma$ as a small perturbation which modifies the evolution of the curvature perturbation. In order words, the equations of motion for the curvature perturbation's mode functions become
\begin{align}\label{eq:curvaturepert}
{\cal R}''+2{\cal H}{\cal R}'+k^2{\cal R}=\frac{\delta {\cal L}^{(2)}_{\rm int}(\tau)}{\delta {\cal R}}\,.
\end{align}
If the right hand side is a small perturbation, we can split the curvature perturbation into ${\cal R}={\cal R}_0+{\cal R}_1$. Then we can solve Eq.~\eqref{eq:curvaturepert} by the Green's function method which yields \cite{Gong:2001he,Joy:2005ep,Gong:2005jr,Bartolo:2013exa,Achucarro:2014msa}
\begin{align}
{\cal R}_1(\tau)=i\int d\eta \left({\cal R}_0(\eta){\cal R}^*_0(\tau)-{\cal R}^*_0(\eta){\cal R}_0(\tau)\right)\frac{\delta {\cal L}^{(2)}_{\rm int}(\eta)}{\delta {\cal R}}\,,
\end{align}
where ${\cal R}_0$ is the solution of the homogeneous equation and it is given in the WKB approximation by
\begin{align}
{\cal R}_0(\tau)=\frac{1}{a(\tau)\sqrt{2k}}e^{-ik\tau}\,.
\end{align}

At this point we can provide a formal estimation of the modulation of the primordial power spectrum, defined by
\begin{align}
P_{\cal R}=\frac{k^3}{2\pi^2}\langle {\cal R}^2\rangle\,.
\end{align}
At the zeroth order we have that $P_{{\cal R},0}=\frac{k^3}{2\pi^2}|{\cal R}_0|^2$ and we obtain the prediction for the primordial spectrum of the scenario at hand. The leading order on ${\cal R}_1$ will depend on whether $\sigma$ develops a background value or not. For instance, if $\sigma$ undergoes classical oscillations we have that
\begin{align}
\frac{\Delta P_{{\cal R},\sigma}}{P_{{\cal R},0}}=\frac{2{\rm Re}[{\cal R}_0^*{\cal R}_1]}{|{\cal R}_0|^2}\,.
\end{align}
Contrariwise, if $\sigma$ has no background value, the leading contribution to ${\cal R}$ comes from a second order perturbative expansion in $\delta\sigma$  ($\delta\sigma$ and ${\cal R}$ are statistically independent at the zeroth order). Thus, we obtain that the quantum fluctuations of $\delta\sigma$ yield a modification given by
\begin{align}
\frac{\Delta P_{{\cal R},\delta\sigma}}{P_{{\cal R},0}}=\frac{|{\cal R}_1|^2}{|{\cal R}_0|^2}\,.
\end{align}
It should be noted that using the in-in formalism one arrives at the same result \cite{Chen:2011zf,Chen:2018cgg}.

We now proceed to provide concrete estimates for the frequency of the resulting oscillatory features in the primordial spectrum in two different mechanisms. First we will review the modulation coming from a background oscillation of $\sigma$ and then by an initial excited quantum state.

\subsection{Classical PSCs in non-inflationary scenarios \label{subsec:bg}}

As mentioned, exciting a massive scalar field classically requires certain sharp feature or initial configuration. Both sharp feature/initial condition and the subsequent oscillation of the massive field have an impact on the density perturbations, generating the sharp feature signal and clock signal, respectively, and the two are smoothly connected. The review of this subsection only focuses on the clock signal. We will comment on this aspect more later.

The Klein-Gordon equation for the background value of an oscillating massive field is given by
\begin{align}
\sigma''+2{\cal H}\sigma'+M^2a^2\sigma=0\,.
\end{align}
Since $aM/{\cal H}>1$, the field $\sigma$, if excited, oscillates around the minimum
\begin{align}
\sigma\sim \sigma_0\left(\frac{a}{a_0}\right)^{-3/2}\left(c_+e^{-i\Omega(\tau)}+c_-e^{i\Omega(\tau)}\right)\,,
\end{align}
where $c_-=c_+^*$ are constants and
\begin{align}\label{eq:Omega}
\Omega=\int aMd\tau=Mt\,.
\end{align}
In the final step of Eq.~\eqref{eq:Omega} we have used that $M$ is constant and that the cosmic time is defined by $dt=ad\tau$. If $M$ was time dependent, we would not have a linear relation between $\Omega$ and $t$.

The background oscillations of $\sigma$ will modulate the evolution of the curvature perturbation ${\cal R}$ and introduce oscillatory features in the power spectrum. A quick evaluation disregarding time dependent amplitudes yields
\begin{align}\label{eq:modulation1}
\frac{\Delta P_{{\cal R},\sigma}}{P_{{\cal R},0}}=\frac{2{\rm Re}[{\cal R}_0^*{\cal R}_1]}{|{\cal R}_0|^2}\sim {\rm Im}\left[\int d\eta e^{-2ik\eta+i\Omega(\eta)}\right]\,,
\end{align}
where we used that ${\cal R}\propto e^{-2ik\tau}$ is in the Bunch-Davies vacuum deep inside the horizon ($k\gg {\cal H}$). The integral \eqref{eq:modulation1} gives it highest contribution at a saddle point where the mode with wavenumber $k$ given by
\begin{align}
2k=\Omega'=Ma
\end{align}
resonates. Evaluating the integral at the saddle point, one roughly obtains that the modulation of the power spectrum reads
\begin{align}\label{eq:DP1}
\frac{\Delta P_{{\cal R},\sigma}}{P_{{\cal R},0}}\sim A\sin\left[\frac{\alpha^2}{|1+\alpha|}\frac{Ma_0}{{\cal H}_0}\left(\frac{2k}{a_0M}\right)^{1+1/\alpha}+\theta\right]\,,
\end{align}
where ${\cal H}_0=\alpha/\tau_0$ and $\theta$ is an arbitrary phase. Note that similar conclusions are obtained directly from the equations of motion. Also, for inflation where $\alpha\sim-1$ the dependence in $k$ in the frequency is logarithmic.

\subsection{Quantum PSCs in non-inflationary scenarios\label{subsec:quantum}}

The situation gets more interesting in alternative scenarios when one considers the quantum fluctuations of the massive field $\delta\sigma$ around its minimum of the potential (see Ref.~\cite{Chen:2018cgg} for more details). In this case, the equations of motion for the mode functions are given by
\begin{align}\label{eq:deltasigma}
\delta\sigma''+2{\cal H}\delta\sigma'+\left(k^2+M^2a^2\right)\delta\sigma=0\,.
\end{align}
The important point is that in contracting scenarios each mode will start in the $M$ dominated regime followed by a $k$ dominated regime, lastly exiting the horizon ${\cal H}$. However, there is a minimum value of $k$ below which the modes will transition from an $M$ dominated regime directly to outside the horizon. This value of $k$ is given by
\begin{align}
k_{\rm min}\equiv aM\left(\frac{aM}{{\cal H}}\right)^{-\frac{\alpha}{1+\alpha}}=a_0M\left(\frac{Ma_0}{{\cal H}_0}\right)^{-\frac{\alpha}{1+\alpha}}\,.
\end{align}
On one hand, the solution of Eq.~\eqref{eq:deltasigma} in the $M$ dominated stage is given by
\begin{align}
\delta\sigma(\tau<\tau_{*})\sim c_{+,M}e^{-i\Omega(\tau)}+c_{-,M}e^{+i\Omega(\tau)}\,,
\label{Eq:nonBD}
\end{align}
where $\tau_{*}$ is the time where $k=aM$. On the other hand, the solution in the $k$ dominated regime reads
\begin{align}
\delta\sigma(\tau>\tau_*)\sim c_{+,k}e^{-ik\tau}+c_{-,k}e^{+ik\tau}\,,
\end{align}
Then, assuming an instantaneous transition and matching at $k=Ma$ we find that
\begin{align}
c_{\pm,k}=c_{\pm,M}\exp\left[\pm i\frac{\alpha^2}{1+\alpha}\left(\frac{k}{k_{\rm min}}\right)^{1+1/\alpha}\right]=c_{\pm,M}\exp\left[\pm i\frac{\alpha^2}{1+\alpha}\frac{Ma_0}{{\cal H}_0}\left(\frac{k}{a_0M}\right)^{1+1/\alpha}\right]\,.
\end{align}
This means that for $k>k_{\rm min}$ the quantum fluctuations $\delta\sigma$ have developed a phase during the transition from the $M$-dominated to $k$-dominated regime. This phase can be encoded in the primordial fluctuations. It should be noted that as in Sec.~\ref{subsec:bg} any time dependence of the mass will modify the induced phase.

Since the field $\delta\sigma$ has no background value, the leading contribution to the curvature perturbation power spectrum is through a second order perturbative expansion, that is
\begin{align}
\frac{\Delta P_{{\cal R},\delta\sigma}}{P_{{\cal R},0}}=\frac{|{\cal R}_1|^2}{|{\cal R}_0|^2}\sim \left|\int d\eta {\cal R}_0(\eta)\delta\sigma(\eta)\right|^2\,.
\end{align}
In this case, it is obvious that there will not be any resonances as $\delta\sigma$ during the $k$-dominated regime oscillates with the same frequency as ${\cal R}$. This time, the frequency of the resulting modulation can be readily estimated by noting that the phase developed by $\delta\sigma$ is time independent. Thus, we can pull it out of the integrals and we have that \cite{Chen:2018cgg}
\begin{align}\label{eq:DP2}
\frac{\Delta P_{\cal R}}{P_{{\cal R},0}} \sim |c_+c_-^*|\sin\left(2\frac{\alpha^2}{1+\alpha}\left(\frac{k}{k_{\rm min}}\right)^{1+1/\alpha}+\theta\right)\sim A\sin\left[2\frac{\alpha^2}{|1+\alpha|}\frac{Ma_0}{{\cal H}_0}\left(\frac{k}{a_0M}\right)^{1+1/\alpha}+\theta\right]\,,
\end{align}
where $\theta$ is an arbitrary phase. Note how in both cases \eqref{eq:DP1} and \eqref{eq:DP2} one obtains an oscillatory modulation with a frequency proportional to $k^{1+1/\alpha}$. This implies that the frequency of these type of oscillatory signals is quite model independent and directly related to the power-law index $\alpha$.

\medskip
\medskip

\begin{figure}
\includegraphics[width=0.5\columnwidth,valign=m]{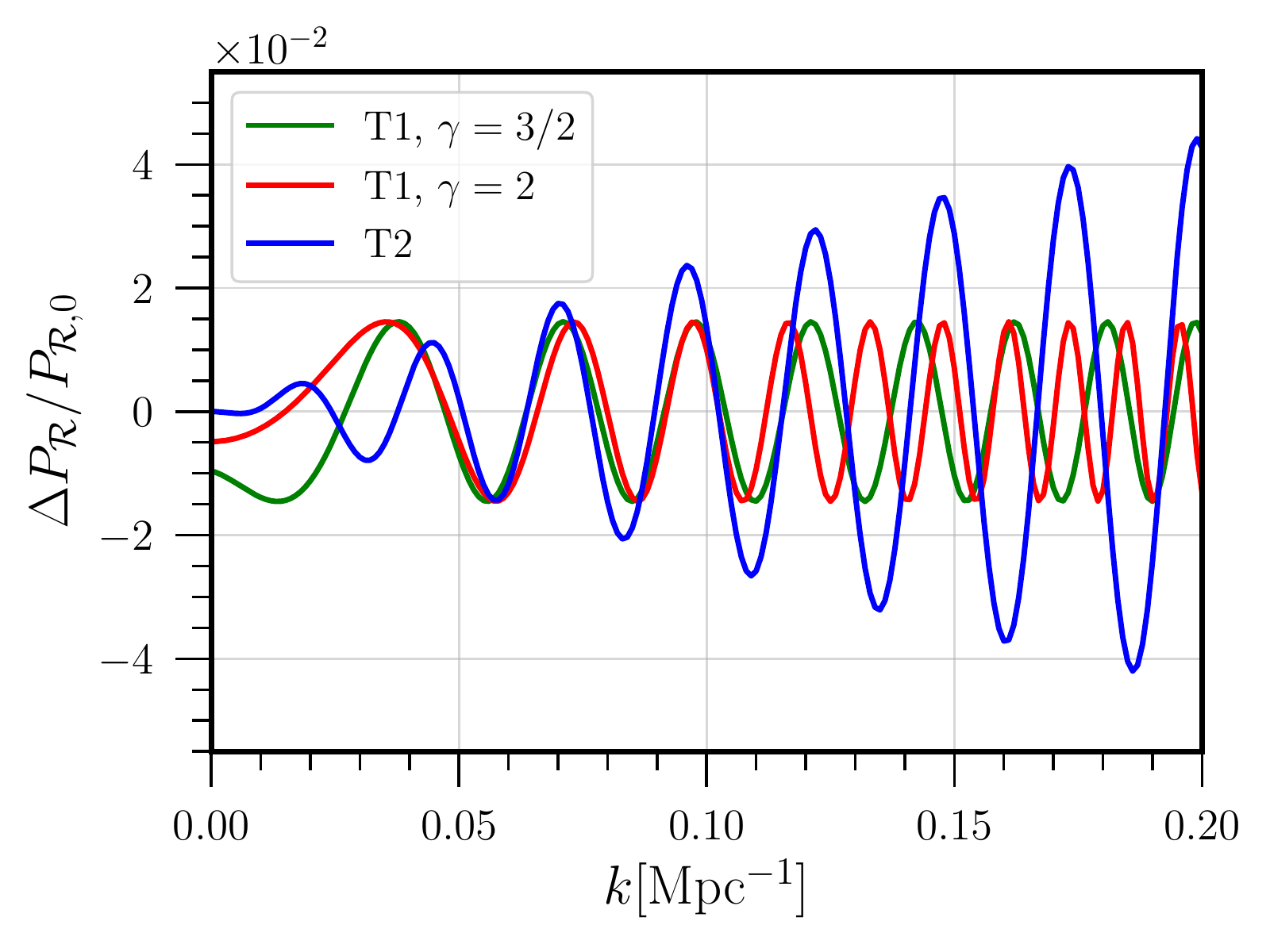}
\caption{ Oscillatory modulation to the primordial power spectrum due to the oscillations of a massive scalar field during a power-law universe Eq.~\eqref{eq:T1}(T1) and a sharp peak in the sound speed of perturbations during inflation Eq.~\eqref{eq:T2}(T2). In green and red we see T1 with $\alpha=2$  ($\gamma=3/2$, $p=2/3$) and $\alpha=1$ ($\gamma=2$, $p=1/2$) respectively with the mean values of table~\ref{Tab:meanT1T2all}. In blue we show T2 with the best fit values of table~\ref{Tab:bestfitT1T2}. Note how they have a similar shape from $k\sim 0.06{\rm Mpc}^{-1}$ to $k\sim 0.11{\rm Mpc}^{-1}$. Since they have a frequency similar to that of the acoustic peaks, they may provide an explanation to the $A_L$ anomaly for $\ell\sim800-1500$. It should be noted that the template T1 (green and red lines) induces a non-zero effect infinitely far from the horizon at $k\to0$ due to the non-Bunch-Davies vacuum assumption in \eqref{Eq:nonBD}. \label{fig:DPs}}
\end{figure}

We make a few comments before proceeding to the data analysis. First, for a massive field to oscillate classically, a sharp feature is necessary in any model to push the field above the bottom of potential. To be complete, the sharp feature also generates a distinctive type of signal in the power spectrum. This signal is sinusoidal and its frequency in $k$-space is approximately constant. In the full signal of classical PSC, the sharp feature signal occupies the larger scales, while the clock signal Eq.~\eqref{eq:DP1} occupies the shorter scales. The two are smoothly connected. Therefore, we point out that the template Eq.~\eqref{eq:DP1} is incomplete and its modification at the larger scales could change the data analysis significantly. An explicit example of the full signal in the inflation case can be found in Ref.~\cite{Chen:2014joa,Chen:2014cwa}. The full signals for alternative-to-inflation models have only been studied schematically \cite{Chen:2011zf,Chen:2011tu} and explicit examples have not be constructed so far.
In contrast, the quantum PSC case reviewed in Sec.~\ref{subsec:quantum} does not suffer from the above problem because quantum fluctuations happen spontaneously in absence of any sharp features.
So, our analysis below mainly applies to the quantum case, and the implications for the classical case should be taken with caution.

Second, from Eq.~\eqref{eq:DP1} we see a direct relation of the $k$ dependence of the frequency with the power-law index $\alpha$. This is due to our constant mass assumption. As mentioned, non-standard clocks, such as time-dependent mass $M$, would modify the direct relation to $\alpha$ \cite{Chen:2011zf,Huang:2016quc,Domenech:2018bnf}. This means that even during inflation one may get more general $k$ dependence in the frequency. This applies to both the classical and quantum PSC cases.
For example, for classical PSCs, Ref.~\cite{Domenech:2018bnf} considered an oscillating spectator massive field during inflation where its mass and the kinetic terms had a coupling to the inflaton. This coupling renders the effective mass time dependent and leads to a $k$ dependent frequency of the oscillatory modulation in the primordial power spectrum. Further study of non-Gaussianities revealed that the $k$ dependence of the amplitude breaks the degeneracy between an oscillatory feature generated during inflation by a massive field with time-dependent mass and an alternative scenario with constant mass. Furthermore, one may argue that a constant mass is a more ``natural'' choice than a fine-tuned interaction during inflation. Also, the number of parameters involved in the case of  time-dependent mass are greater than in the constant mass which may cause the model to be less favored by Occam's razor.
In this paper, we study the constant mass case. We leave the analysis and distinction between time-independent and time-dependent cases for future studies.

Third, in absence of any massive fields, sharp feature alone, such as that in the potential or sound speed, also generates oscillatory signals in the power spectrum. Actually, this is the first type of oscillatory feature signals that have been studied extensively in the literature \cite{Starobinsky:1992ts,Wang:1999vf,Adams:2001vc,Chen:2006xjb,Adshead:2011jq, Bean:2008na,Chen:2008wn,Achucarro:2010da, Miranda:2012rm, Fergusson:2014hya, Fergusson:2014tza, Gao:2012uq}.
The model-independent feature of this type of signals is the sinusoidal running with approximately constant frequency, generically described by the following template \cite{Chen:2008wn,Fergusson:2014hya,Fergusson:2014tza}:
\begin{align}
  \frac{\Delta P_{\cal R}}{P_{{\cal R},0}} =
    C \sin \left( \frac{2 k}{k_0} + \theta \right) ~,
    \label{Eq:sharpfeature}
\end{align}
where $k_0$ is a constant and we have ignored a $k$-dependent envelop whose form is highly model-dependent. Here we emphasize that, unlike the PSC signals, the above qualitative behavior of the sharp feature signal, obtained in the inflationary scenario, is in fact the same even in alternative scenarios to inflation \cite{Chen:2011zf}.
Notice that the template (\ref{Eq:sharpfeature}) naively corresponds to the case $p=1$ ($\alpha=\infty$) in (\ref{eq:DP2}). This makes sense because there is no primordial scenario that has $p=1$, in which case the wavelengths of quantum fluctuations expand or contract at the same speed as the horizon size and would never exit the horizon.
More precise templates from specific types of sharp features are also used in data analysis. In the following analyses, we will use the template derived in Ref.~\cite{Achucarro:2014msa,Domenech:2019cyh}, to be shown explicitly in Sec.~\ref{sec:results}, for a specific type of sharp feature in the inflation model.

\section{CMB constraints and lensing anomaly\label{sec:results}}

In this section we present the CMB constraints on the aforementioned oscillatory features in the primordial power spectrum. We also include a constant rescaling of the lensing spectra parametrized by the $A_L$ parameter \cite{Calabrese:2008rt}, where $A_L=1$ for $\Lambda$CDM. For the primordial power spectrum we assume as usual that it is of a power-law form, namely
\begin{align}
P_{{\cal R},0}=A_s\left(\frac{k}{k_{\rm pivot}}\right)^{n_s-1}
\end{align}
where $k_{\rm pivot}=0.05{\rm Mpc}^{-1}$ and $A_s$ and $n_s$ are free parameters to fit with the data. Below we present the explicit form of the modulations as well as their relation to the lensing anomaly. Later, we describe the methodology and discuss the main results.

\paragraph{Templates for the oscillatory modulation:} On top of the power-law ansatz, we consider two templates for the oscillatory modulation. First, denoting as template 1 (T1), we use the oscillatory feature from an oscillating massive field in a power-law universe, which is given by
\begin{align}\label{eq:T1}
\frac{\Delta P_{{\cal R},T1}}{P_{{\cal R},0}}\approx A\sin\left[\omega\left(\frac{k}{k_*}\right)^{\gamma}+\phi\,\pi\right]\,,
\end{align}
where $A$, $\omega$, $\gamma$ and $\phi$ are free parameters and we fix $k_*=0.163 {\rm Mpc}^{-1}$. Second, denoting as template 2 (T2), we use the oscillatory feature resulting from a sharp peak in the sound speed of pertrubations during inflation, that is \cite{Achucarro:2014msa,Domenech:2019cyh}
\begin{align}\label{eq:T2}
\frac{\Delta P_{{\cal R},T2}}{P_{{\cal R},0}}=B\sqrt{\pi}\frac{k\eta_f}{b_s}{\rm e}^{-\frac{k^2\eta_f^2}{b_s^2}}\left\{\sin(2k\eta_f)+\frac{\cos(2k\eta_f)}{k\eta_f}-\frac{1}{2}\frac{\sin(2k\eta_f)}{(k\eta_f)^2}\right\}\,,
\end{align}
where $B$, $b_s$ and $\eta_f$ are free parameters which respectively indicate the amplitude, the sharpness and the position of the feature during inflation.\footnote{The feature is assumed to be of the form
$ c_s^2=1+B{\rm e}^{-b_s^2\ln^2\left[\tau/\eta_f\right]}$,
where $B<0$ to ensure subluminal propagation.  We will consider $B\sim {\cal O}(0.01)$.  For our purpose, the amplitude of non-Gaussianity $f_{NL}\sim -0.27\,B b_s^2$  \cite{Achucarro:2014msa,Domenech:2019cyh,Akrami:2019izv} does not impose severe constraints on the model. The constraints by the Planck data on non-Gaussianities with oscillatory scale-dependence are relatively loose, comparing to other forms of non-Gaussianities.} \\

\begin{table}
\begin{center}
\begin{tabular}{|| c | c | c | c | c | c | c ||}
    \hline
   \begin{tabular}[c]{@{}c@{}} Model (TTTEEE \\ 68\% confidence) \end{tabular}  &  $A_L$ &  $A$  & $\gamma$  & $\omega$ &$\phi$& $\chi^2$ 
    \\ [0.5ex]
    \hline\hline
    $\Lambda CDM$  & - & - & - & - & - & $2767$
     \\[0.5ex]
    $\Lambda CDM$ + $A_L$ & $1.18_{-0.07}^{+0.07}$ &  - & - & - & - & $2760$
     \\[0.5ex]
    $\Lambda CDM$ + $A_L$ + T1 & $1.14_{-0.07}^{+0.07}$ & $0.012_{-0.005}^{+0.005}$ & $1.56_{-0.38}^{+0.18}$ & $36_{-3}^{+3}$ & $-0.8_{-0.5}^{+0.6}$ & $2752$
    \\[0.5ex]
      $\Lambda CDM$ + T1 & - & $0.015_{-0.005}^{+0.005}$ & $1.69_{-0.25}^{+0.25}$ & $37_{-3}^{+3}$ & $-0.5_{-0.3}^{+0.5}$ & $2754$
    \\[0.5ex]
  \hline
\end{tabular}
\end{center}
\begin{center}
\begin{tabular}{|| c | c | c | c | c | c ||}
    \hline
   \begin{tabular}[c]{@{}c@{}}  Model (TTTEEE \\ 68\% confidence)\end{tabular} &  $A_L$ &  $B$  & $\eta_f$ &$b_s$ & $\chi^2$ 
    \\ [0.5ex]
    \hline\hline
  $\Lambda CDM$ + $A_L$ + T2 & $1.14_{-0.07}^{+0.07}$ & $-0.05_{-0.03}^{+0.03}$  & $124_{-3}^{+3}$ & $71_{-10}^{+29}$ & $2756$ \\ [0.5ex]

  $\Lambda CDM$ + T2 & - & $-0.07_{-0.02}^{+0.04}$ & $123_{-2}^{+2}$ & $70_{-10}^{+30}$ & $2759$ \\ [0.5ex]
    \hline
\end{tabular}
\end{center}
    \caption{Constraints on  T1 (top) and T2 (bottom) using the full TTTEEE + low E CMB likelihood from Planck 2018 \cite{Aghanim:2018eyx}. For comparison we included the constraints on the $A_L$ parameter with and without T1 or T2. In both cases we see that while $\Lambda$CDM +$A_L$ yields $A_L\neq 1$ slightly above $2\sigma$, $\Lambda$CDM +$A_L$ +T1 reduces the $A_L$ tension to slightly below $2\sigma$. On the top table, we also note that $\Lambda$CDM +T1 has a $\Delta\chi^2=-13$ and $\Delta\chi^2=-6$ in comparison respectively with $\Lambda$CDM and $\Lambda$CDM +$A_L$. Also we find that in the $\Lambda$CDM + T1 the amplitude of the oscillation $A=0$ as well as the value $\gamma=1$ are excluded close to $3\sigma$. For the bottom table, the value of $\chi^2$ of $\Lambda$CDM +T2 is not significantly lower than $\Lambda$CDM +$A_L$ and is $\Delta\chi^2=-8$ with respect to $\Lambda$CDM. Also we find that the amplitude of the oscillation $B=0$ is allowed within $3\sigma$. \label{Tab:meanT1T2}}
\end{table}

\paragraph{Possible relation to the lensing anomaly:} The lensing anomaly appears when the lensing spectra is allowed to be rescaled by a constant factor $A_L$ \cite{Calabrese:2008rt}, where $A_L=1$ corresponds to $\Lambda$CDM. It is essentially used as a diagnostic of the Planck data \cite{Aghanim:2018eyx} and, therefore, carries no physical meaning by itself. Interestingly, Planck reported that there seems to be $10\%$ more lensing in the smoothing of the power spectrum than expected, especially around $1100\lesssim\ell \lesssim  2000$, with $A_L\sim 1.18$ and in tension with $A_L=1$ at $2.8\sigma$. It should be noted that including the lensing likelihood $A_L$ is compatible with $A_L=1$ within $2\sigma$ \cite{Aghanim:2018eyx}. This means that the preference for the anomaly is only due to the power spectrum. Thus, we will focus on the anomaly in the temperature and polarization power spectra.

The effect of lensing on the CMB power spectrum is to smooth the structure of the acoustic peaks. This could be plausibly explained by an oscillatory modulation of the primordial power spectrum with a similar frequency of the acoustic peaks but out of phase (at least around $1100\lesssim\ell \lesssim  2000$). In this way, it would naively reduce the size of the peaks and fill in the troughs. Although in practice the oscillatory modulation does not completely mimic the smoothing effect of lensing \cite{Domenech:2019cyh}, it may be enough to reduce the tension.  In addition, it is known that oscillatory modulations and the effect of lensing are degenerate \cite{Hazra:2014jwa,Hazra:2018opk}. This could imply that the extra lensing observed may hide an oscillatory modulation of the primordial spectrum. It may present a power-law structure like T1 \eqref{eq:T1} with the power-law index of $\gamma\sim O(1)$, and the case T2 \eqref{eq:T2} with $\gamma=1$. Thus, the lensing anomaly could potentially hide a fingerprint of alternative scenarios to inflation.

\paragraph{Methodology:}

We use the Boltzmann system solver \texttt{CLASS}\footnote{\url{https://github.com/lesgourg/class_public}} \cite{Blas:2011rf} modified to include the $A_L$ rescaling of the lensing spectra to calculate the resulting CMB anisotropies. We do the Bayesian exploration of the parameter space with the MonteCarlo sampler \texttt{MontePython}\footnote{\url{https://github.com/brinckmann/montepython_public}} \cite{Audren:2012wb,Brinckmann:2018cvx}. Furthermore, we use the full TTTEEEE+low E(+low T) CMB likelihood from Planck 2018 \cite{Aghanim:2018eyx} and all results will refer to TTTEEE unless stated otherwise. In the beginning, we do not include the lensing likelihood since the anomaly is driven mainly by the power spectra. We later add the lensing likelihood to the candidate with the lowest $\chi^2$ to see if it has any substantial impact on the parameters.  We also consider the TT+low E(+low T) CMB likelihood to study the impact of polarization on the value of the power-law index $\gamma$ of T1 \eqref{eq:T1}. We restrict ourselves to values of the frequency of T1 \eqref{eq:T1} and T2 \eqref{eq:T2} relevant to the $A_L$ anomaly, i.e. similar to the acoustic peaks, which from Refs.~\cite{Aghanim:2018eyx,Domenech:2019cyh} is $\omega\sim 40$ and $\eta_f\sim140$. This is to prevent multimodal posteriors, i.e. likelihoods with peaks nearby, which is an interesting challenge for data analysis but it is out of the scope of this paper.  
For this reason, we only studied the first relevant peak by restricting the boundaries of the priors. Additionally, a comparison by eye of the oscillatory modulation with the shape of the acoustic peaks reveals that $\gamma\sim O(1)$. Thus, we focus on the power-law index $\gamma$ to be $0.05\lesssim\gamma\lesssim 5$. For $\gamma>5$ the oscillatory modulation barely fits the residuals over one acoustic peak. We leave the search with wider parameter space and multimodality for future studies.

Before proceeding to the results, we note two important differences with respect to the analysis by the Planck team \cite{Aghanim:2018eyx}. First, in this paper, we obtained $\chi^2$ values directly by the likelihood with the MCMC analysis without the use of further samplers like BOBYQA \cite{powell2009bobyqa}. Therefore, the values of $\chi^2$ might be slightly different than those of Planck. However, the MCMC best fit parameters and the contours do not change much if the actual $\chi^2$ of the global best fit is a bit different. Second, the modification of the CLASS code to include the $A_L$ anomaly only modulates the lensing effect in the temperature and polarization power spectrum, i.e. it only modifies the lensing spectrum that appears in the transfer functions. It does not rescale the trispectrum and, therefore, it does not rescale the lensing spectrum that is contrasted using the lensing likelihood. Contrariwise, in the Planck analysis \cite{Ade:2013zuv,Aghanim:2018eyx} they define $A_L$ such that it rescales the trispectrum as well. Our implementation of $A_L$ is therefore different and corresponds to including the rescaling of the trispectrum $A_{\phi\phi}$ \cite{Ade:2013zuv} such that $A_{\phi\phi}\times A_L=1$. For this reason, the derived value of $A_L$ when we include the lensing likelihood will be slightly higher than that of Planck. Nevertheless, this does not change the main conclusions of this paper as it only concerns the bounds on $A_L$ including the lensing likelihood. 

\begin{figure}
\begin{tabular}{cccc}
\includegraphics[width=0.34\columnwidth,valign=m]{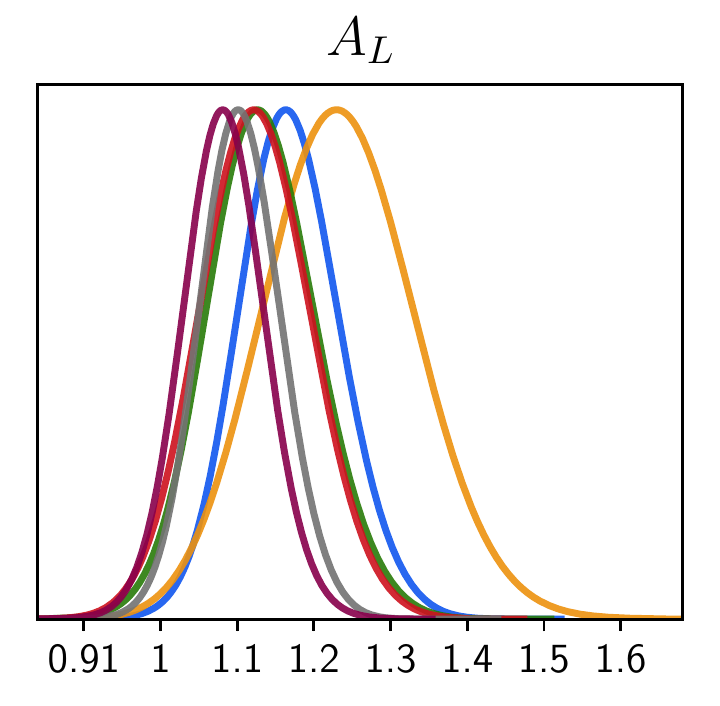}&
\hspace{-15mm}
\includegraphics[width=0.22\columnwidth,valign=m]{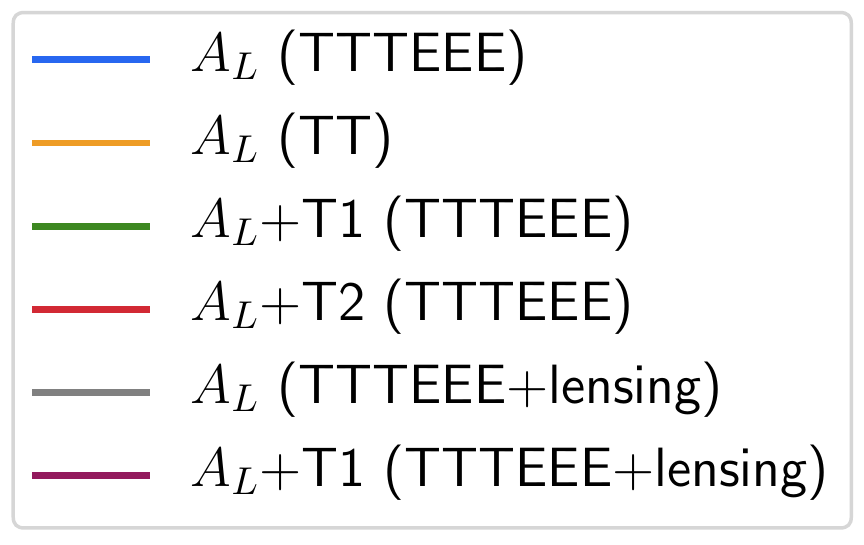}&
\hfill
\includegraphics[width=0.34\columnwidth,valign=m]{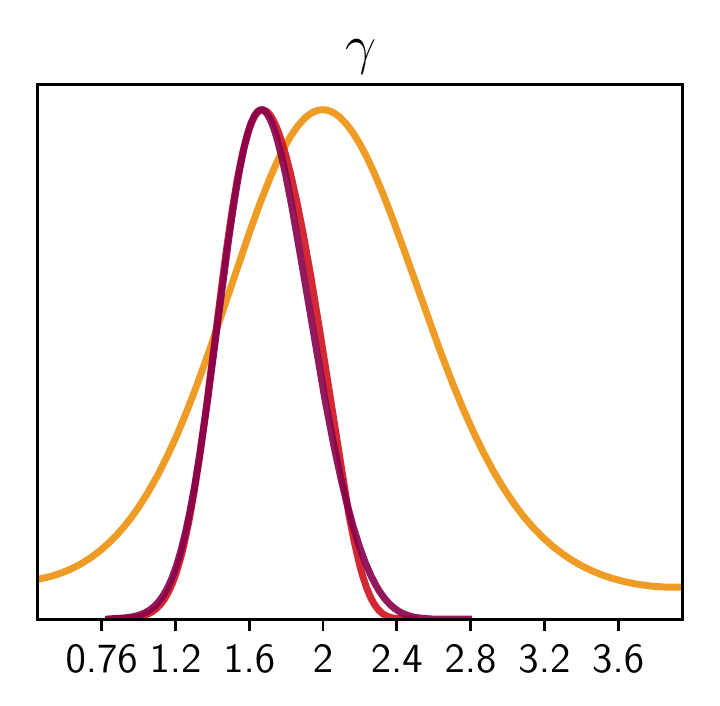}
&
\hspace{-15mm}
\includegraphics[width=0.2\columnwidth,valign=m]{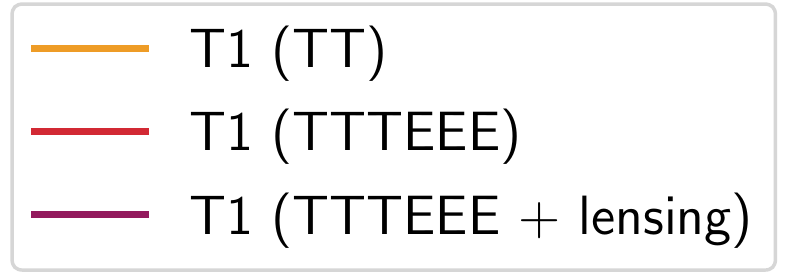}

\end{tabular}
\caption{On the left we show the 1D marginalized likelihood for the $A_L$ parameter using the full TTTEEE + low E Planck likelihood respectively for $A_L$ (blue), $A_L$+T1 \eqref{eq:T1} (green) and $A_L$+T2 \eqref{eq:T2} (red) cases as well as the TT + low E Planck likelihood only for $A_L$. We also show the TTTEEE + low E + lensing Planck likelihood for $A_L$ (grey) and $A_L$+T1 (purple). First see how including polarization lowers the preferred value of $A_L$ but in $\Lambda CDM$ it is still $A_L\neq 1$ at $2.8\sigma$. Second, note how including the modulations T1 or T2 reduces the preferred value of $A_L$ reducing the tension a bit below $2\sigma$. If we include lensing then the statistical significance of the anomaly goes a bit below $2\sigma$ for both $A_L$ and $A_L$+T1, the latter with the lowest value for $A_L$.
On the right we present the 1D marginalized likelihood for the power-law index $\gamma=1+1/\alpha$ in the modulation T1 \eqref{eq:T1} using the TTTEEE+low E (red), TTTEEE+low E+lensing (purple) and TT+low E (orange) Planck likelihoods.
Note that TT+low E prefers a value $\gamma\sim 2$.
Including polarization and lensing lowers the value of $\gamma$ from $\gamma\sim 2$ to $\gamma \sim 1.7$. Note that $\gamma=1$  is excluded close to $3\sigma$. See tables~\ref{Tab:meanT1T2} and \ref{Tab:meanT1lensing} for details on the $1\sigma$ constraints. \label{fig:ALgamma}}
\end{figure}

\subsection{Results\label{subsec:results}}

\begin{figure}
\includegraphics[width=0.34\columnwidth]{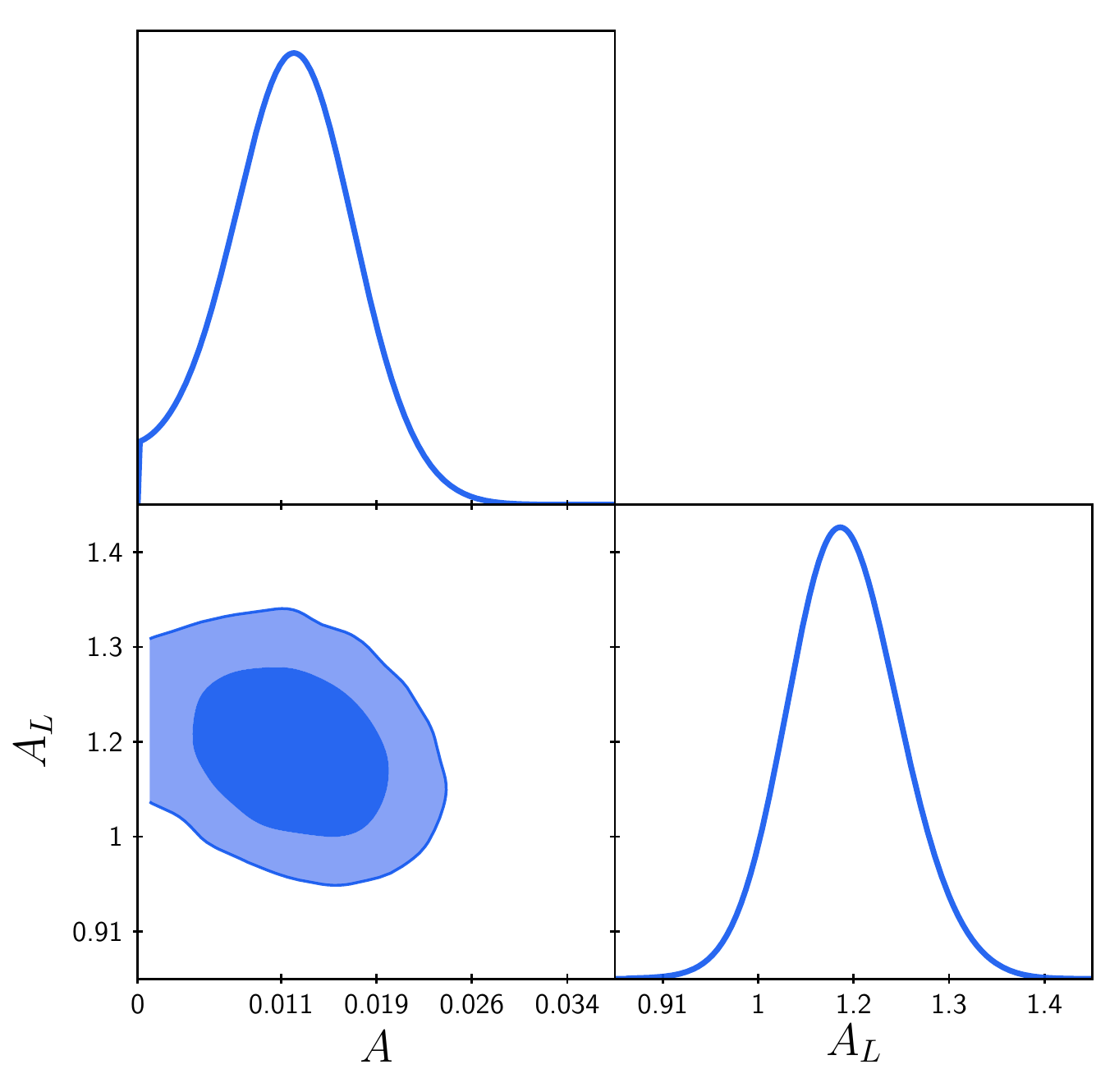}
\hspace{2cm}
\includegraphics[width=0.34\columnwidth]{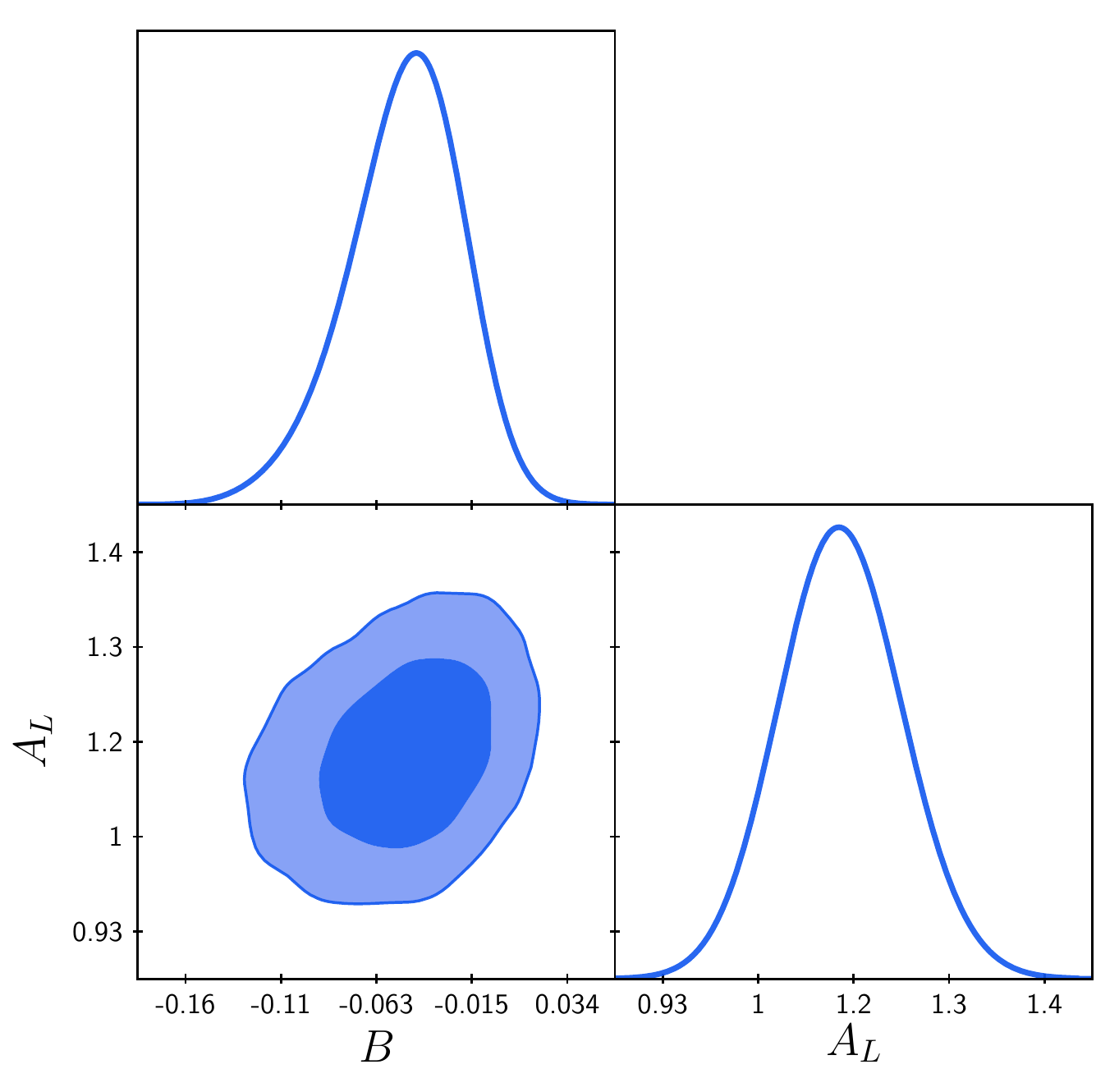}
\caption{Correlation between the $A_L$ parameter and the amplitude of the oscillatory feature in the primordial spectrum from T1 (left) and T2 (right). From the figure the degeneracy between $A_L$ and the oscillatory feature is clear. The larger the amplitude $A$ (or $B$) the smaller the $A_L$ parameter. However, the $A_L$ anomaly persists at $2\sigma$. This is in contrast with the results of Ref.~\cite{Akrami:2018odb} Fig.~(29), where they use very fine-tuned oscillatory feature with constant frequency and Gaussian-envelope to remove the $A_L$ tension. \label{fig:correlation}}
\end{figure}

\begin{figure}[!htp]
\includegraphics[width=0.60\columnwidth,valign=m]{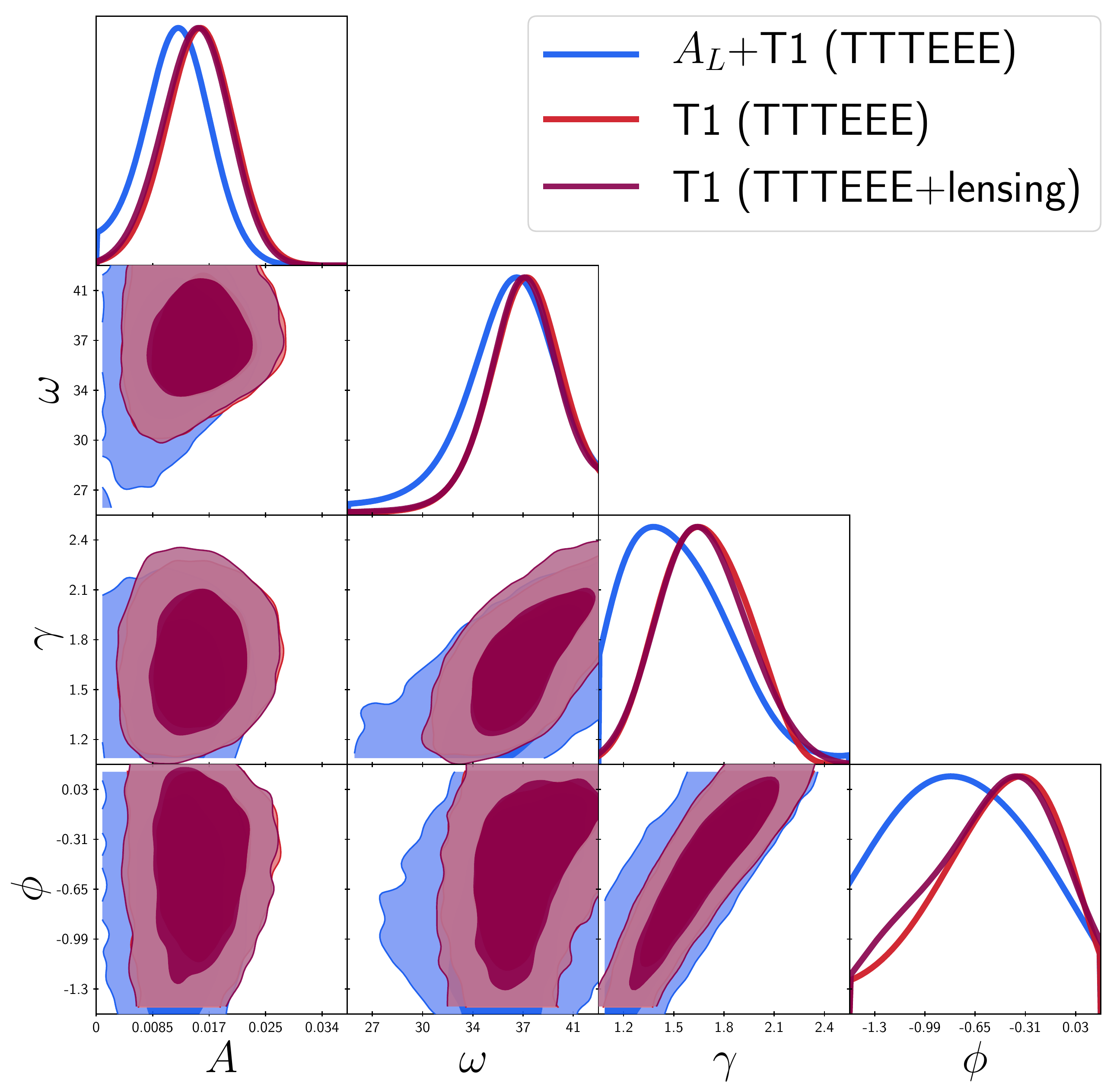}
\includegraphics[width=0.50\columnwidth,valign=m]{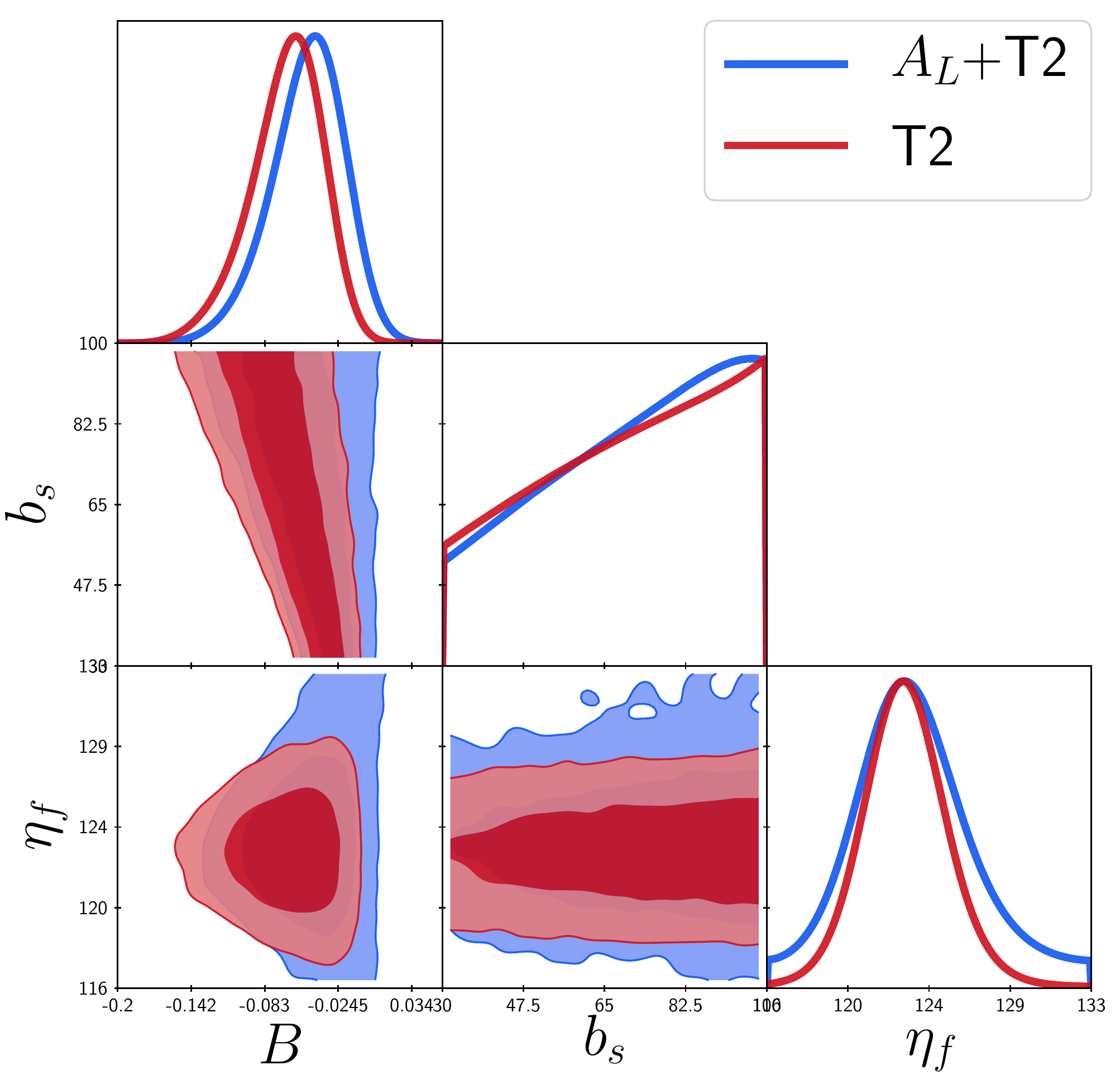}
\caption{1D and 2D marginalized likelihoods with the $1\sigma$ and $2\sigma$ contours using the full TTTEEE+low E Planck likelihoods. On the top figure, we show $A_L$+T1 (blue) and T1 (red) \eqref{eq:T1}. Note that $\omega$ and $\gamma$ show a mild degeneracy, where an increase of $\gamma$ also leads to increase of $\omega$. This might go against the expectations that in order to fit the frequency of the acoustic peaks an increase of $\gamma$ should be compensated by a decrease of $\omega$. Our result is due to a pivot scale $k_*$ which is too large. If we choose a pivot scale relevant to the lensing anomaly, e.g. $k\sim 0.1{\rm Mpc}^{-1}$, the degeneracy has a negative slope as expected. On the bottom figure, we have $A_L$+T2 (blue) and T2 (red) \eqref{eq:T2}. We restricted our study to the first relevant peak near the frequency of the acoustic peaks due to multimodality. \label{fig:1d2dT1T2}}
\end{figure}

The main results are shown in tables~\ref{Tab:meanT1T2} and \ref{Tab:meanT1lensing} and Figs.~\ref{fig:ALgamma} and \ref{fig:1d2dT1T2}. We divided the discussion between the implications to the lensing anomaly and alternative scenarios to inflation. Before that, we proceed to briefly describe the totality of our results. First, in table~\ref{Tab:meanT1T2} we present the CMB constraints respectively on T1 and T2 with the full TTTEEE+low E likelihood at $68\%$ confidence including their $\chi^2$ values. Later, in table ~\ref{Tab:meanT1lensing} we show the results on T1 using the TTTEEE+low E+lensing likelihood. Second, in Fig.~\ref{fig:ALgamma} we show the 1D marginalized likelihood for the $A_L$ parameter for T1 and T2, including the Planck results for TTTEEE, TTTEEE+lensing and TT, and the power-law index of the frequency $\gamma$.
Third, Fig.~\ref{fig:correlation} shows the correlation between the $A_L$ parameter and the amplitude of the oscillatory feature T1 and T2 and Fig.~\ref{fig:1d2dT1T2} shows the 1D and 2D marginalized likelihoods with the $1\sigma$ and $2\sigma$ contours respectively for T1 and T2. To illustrate our results we plotted the residuals for the best fit values of T1 and T2 in Fig.~\ref{fig:T1T2TE}, respectively given in table \ref{Tab:bestfitT1T2}. In the appendix we studied the model T1 for fixed values of the power-law index $\gamma$ and the constraints at $68\%$ confidence with TTTEEE+low E are written in table \ref{Tab:meanT1T2all}. We provide the best fit values of the model T1 including cosmological parameters in table \ref{Tab:bestfitall}. Also in the appendix we include the case of T1 using the TT+low E Planck likelihood to show the dependence of $\gamma$ on the polarization. For this case, the constraints at $68\%$ confidence and the best fit values are given in table \ref{Tab:allTT} and the 1D and 2D marginalized likelihoods in Fig.~\ref{fig:1d2dT1TT}. Finally, an illustration of the best fit model for TT+low E residuals is shown in Fig.~\ref{fig:T1TT}.

\paragraph{Implications for the lensing anomaly:} Let us first focus on the CMB constraints to the $A_L$ parameter of table \ref{Tab:meanT1T2}. Using the TTTEEE+low E data we find that at $68\%$ confidence level
\begin{align}
A_L=\left\{
\begin{aligned}
&1.18_{-0.07}^{+0.07}\qquad (\Lambda{\rm CDM}\,+\,A_L )\,,\\
&1.14_{-0.07}^{+0.07}\qquad (\Lambda{\rm CDM}\,+\,A_L\,+\,{\rm T1})\,.
\end{aligned}
\right.
\end{align}
Similarly, we obtain $A_L=1.14_{-0.08}^{+0.07}$ for $\Lambda$CDM+$A_L$+T2. If we use TT alone we find that $A_L=1.25_{-0.10}^{+0.10}$. On one hand, for $\Lambda$CDM+$A_L$ we recover Planck's results on the lensing anomaly with a statistical significance of $2.8\sigma$. Including lensing further reduces the tension to $2\sigma$ with $A_L=1.12^{+0.05}_{-0.06}$ for $\Lambda$CDM+$A_L$. This value is different from Planck's results $A_L=1.07\pm0.04$ \cite{Aghanim:2018eyx} due to the fact that our CLASS code does not rescale the trispectrum (for details see the discussion before Sec.~\ref{subsec:results}). Thus, the smaller value of $A_L$ is only due to a reduction of the parameter degeneracy.  On the other hand, if we include T1 and T2 we reduce the tension to slightly below $2\sigma$. Including lensing we obtain $A_L=1.10^{+0.05}_{-0.06}$ for $\Lambda$CDM+$A_L$+T1 with $A_L=1$ compatible at $1.7\sigma$. As expected, the lensing anomaly is sensitive to oscillatory modulations in the primordial spectrum (see Fig.~\ref{fig:correlation} for the degeneracy between the amplitude of the oscillatory feature and $A_L$). It should be noted that although the $A_L$ anomaly is not significantly eased, $(i)$ it becomes less statistically significant and $(ii)$ the amplitude of the modulations is non-vanishing within $3\sigma$ for T1 and $2\sigma$ for T2.
These results are better illustrated in the plot at the left of Fig.~\ref{fig:ALgamma} where we show the likelihood for the $A_L$ parameter in all the different cases.

\begin{figure}[H]
\includegraphics[width=0.49\columnwidth,valign=m]{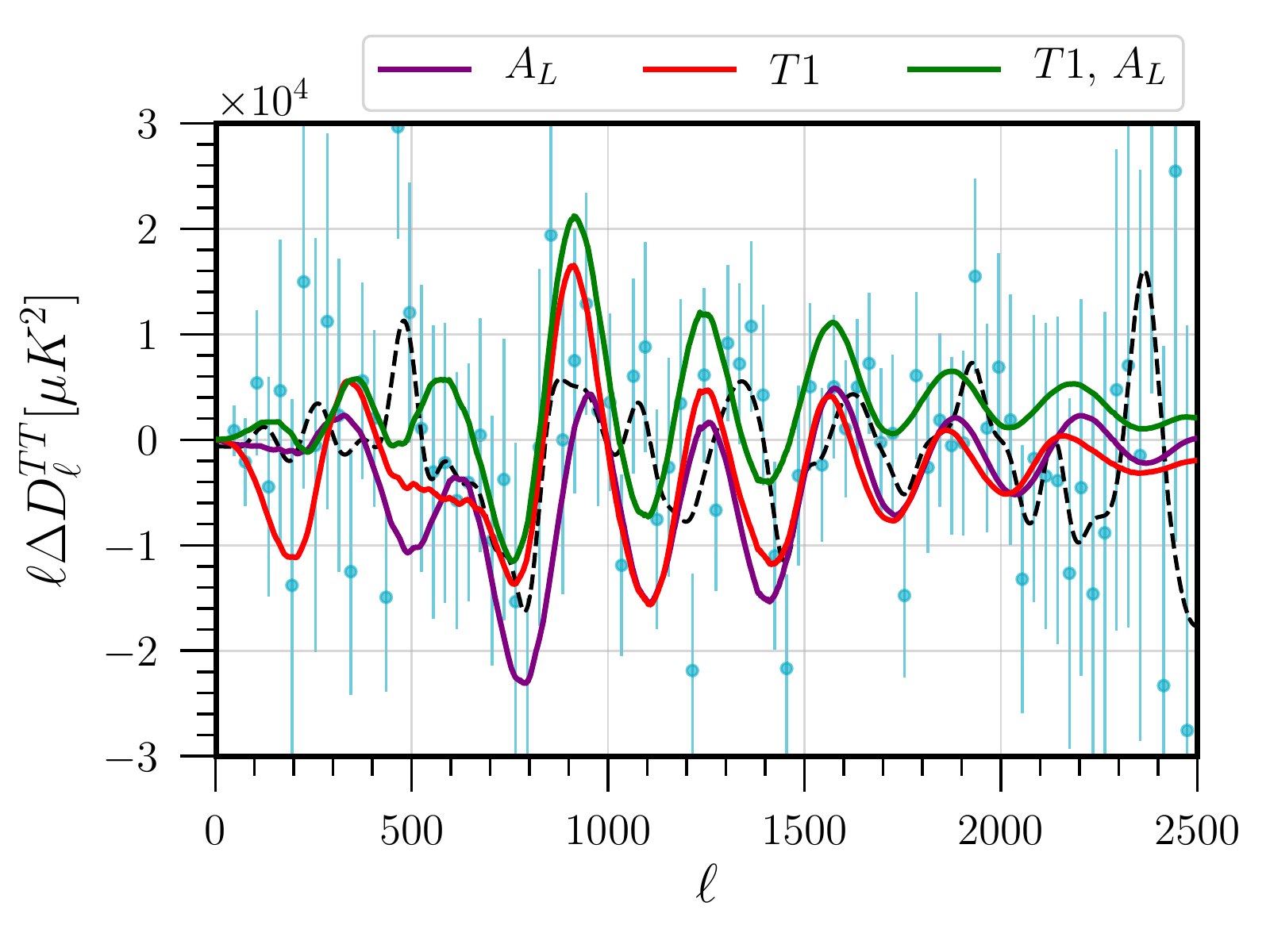}
\includegraphics[width=0.49\columnwidth,valign=m]{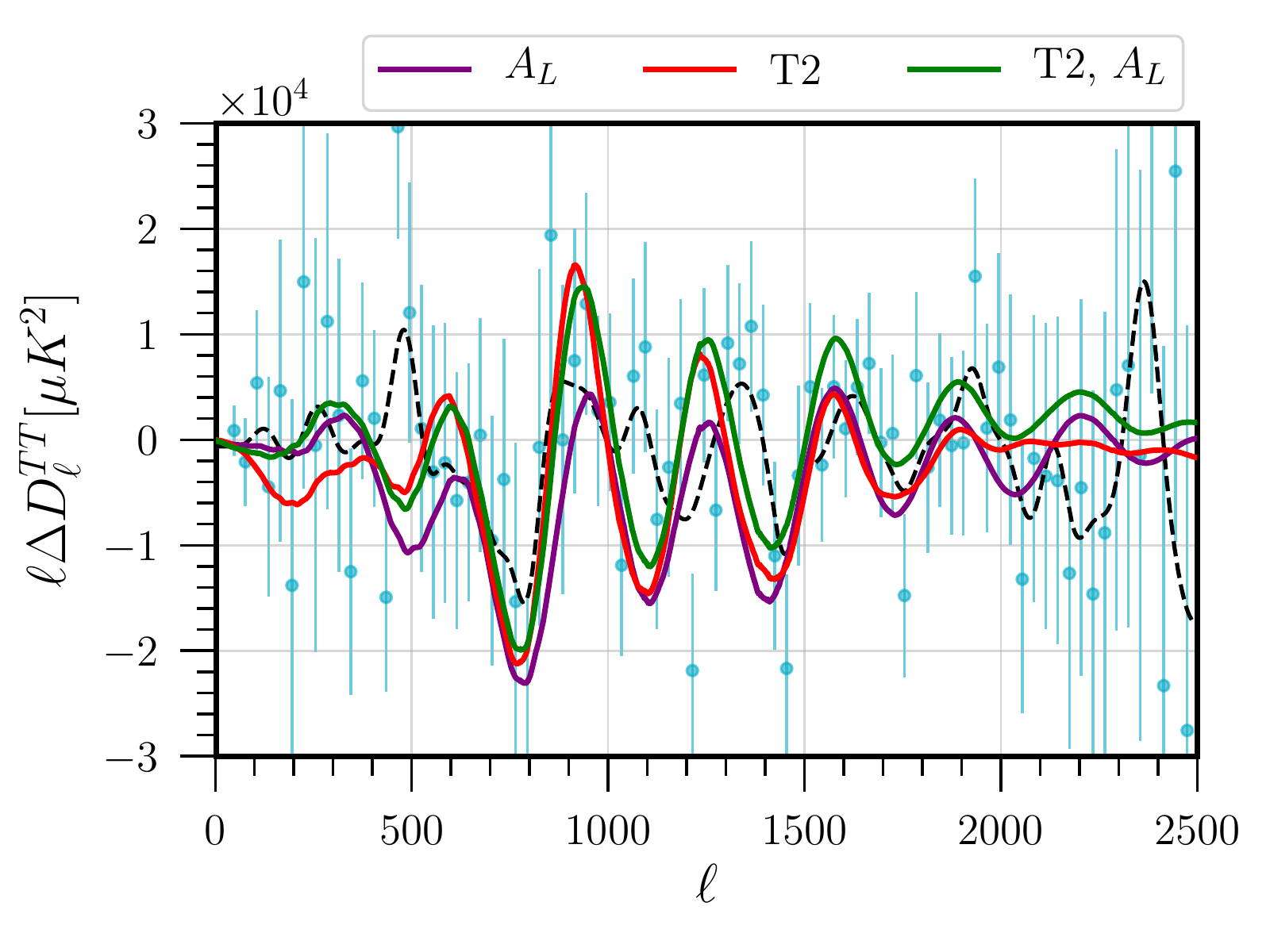}\\
\includegraphics[width=0.49\columnwidth,valign=m]{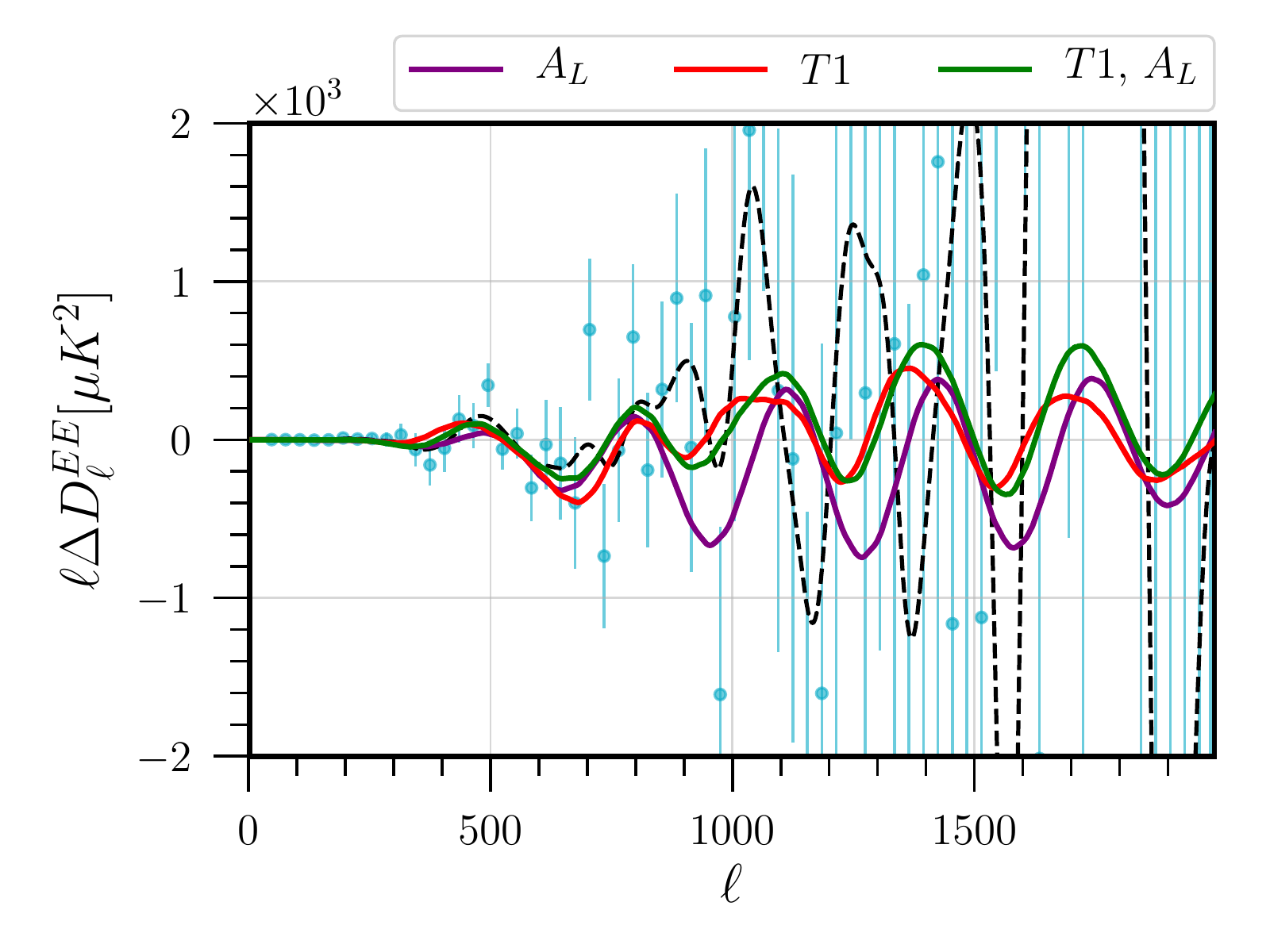}
\includegraphics[width=0.49\columnwidth,valign=m]{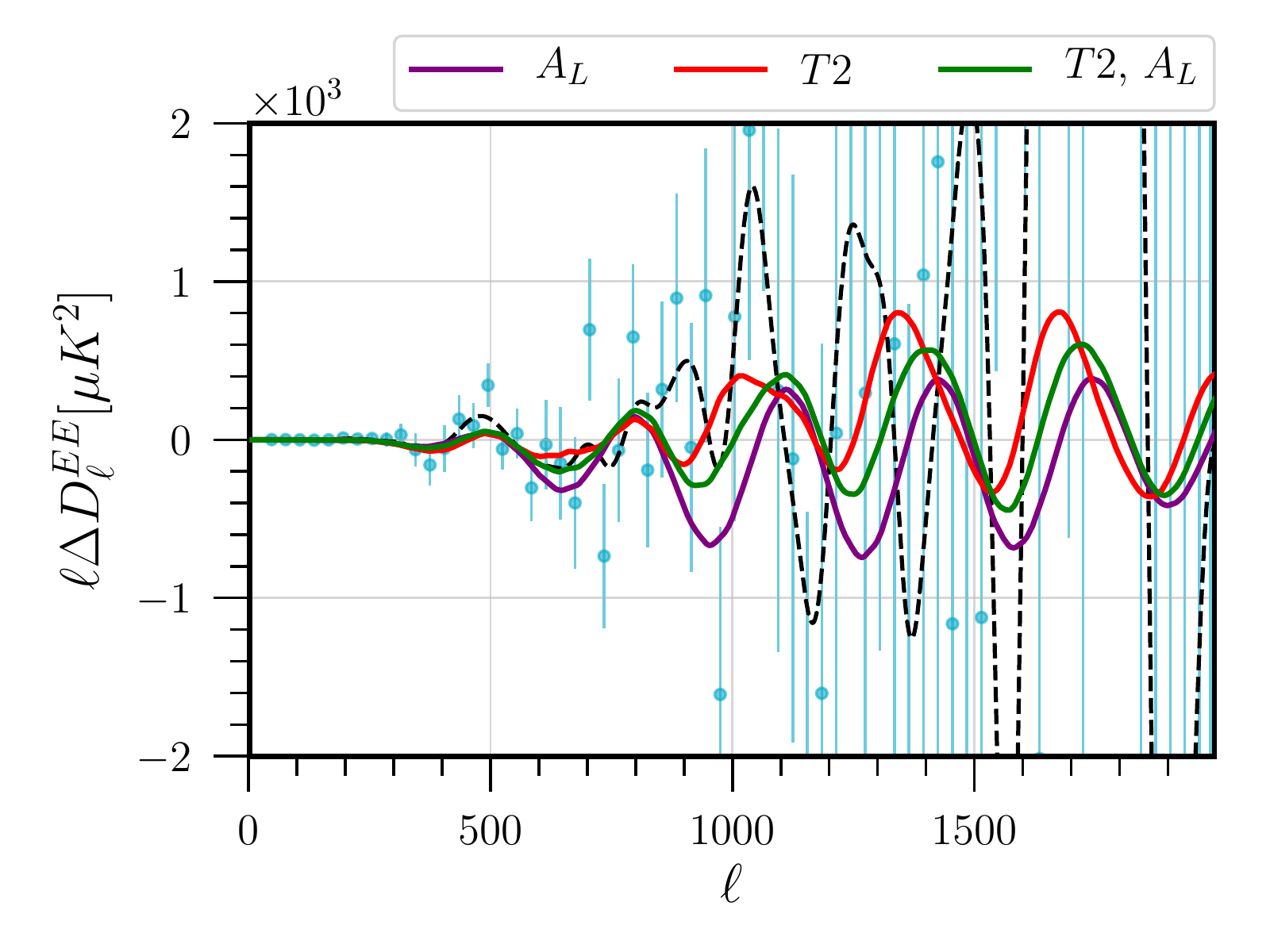}\\
\includegraphics[width=0.49\columnwidth,valign=m]{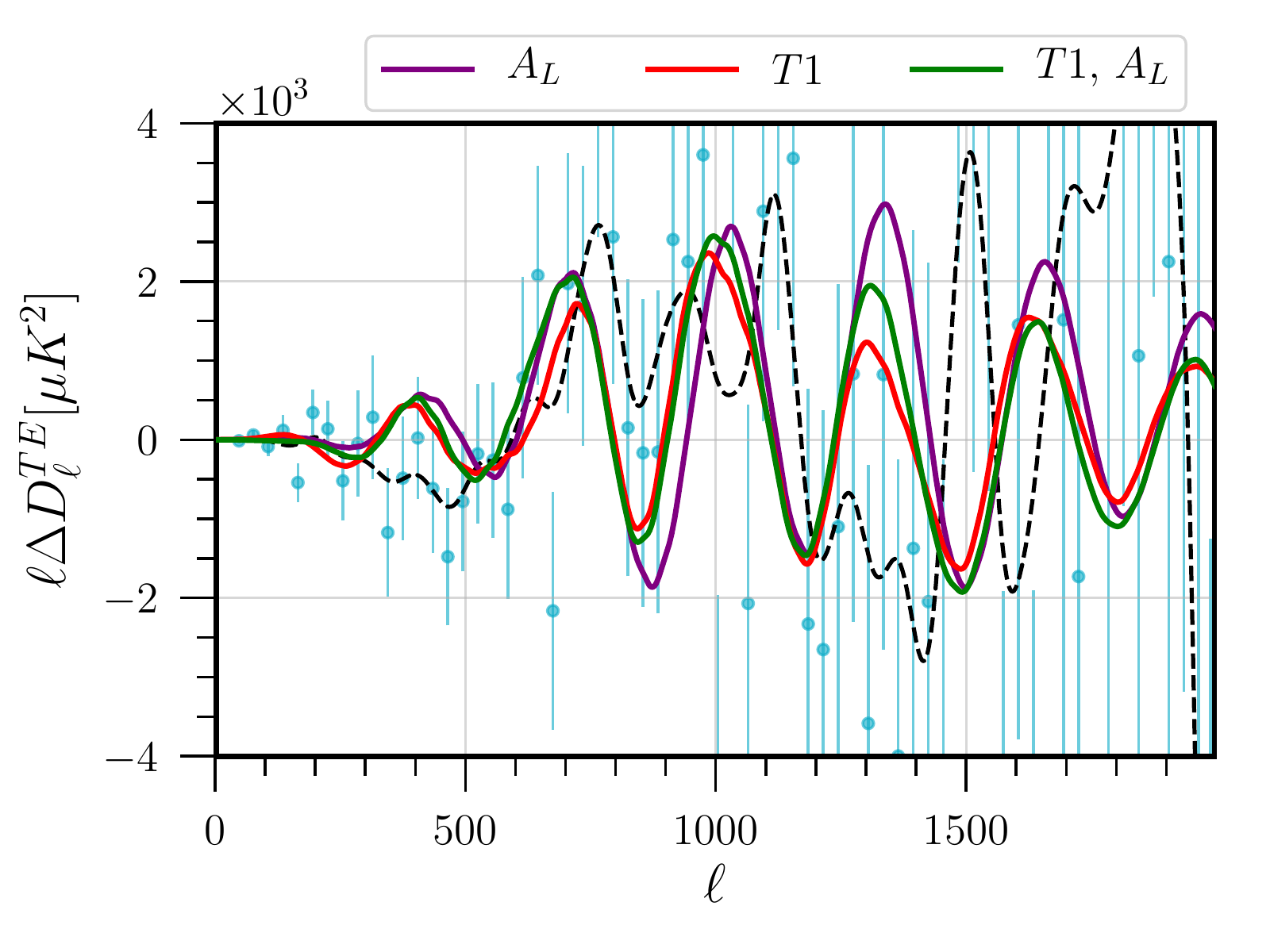}
\includegraphics[width=0.49\columnwidth,valign=m]{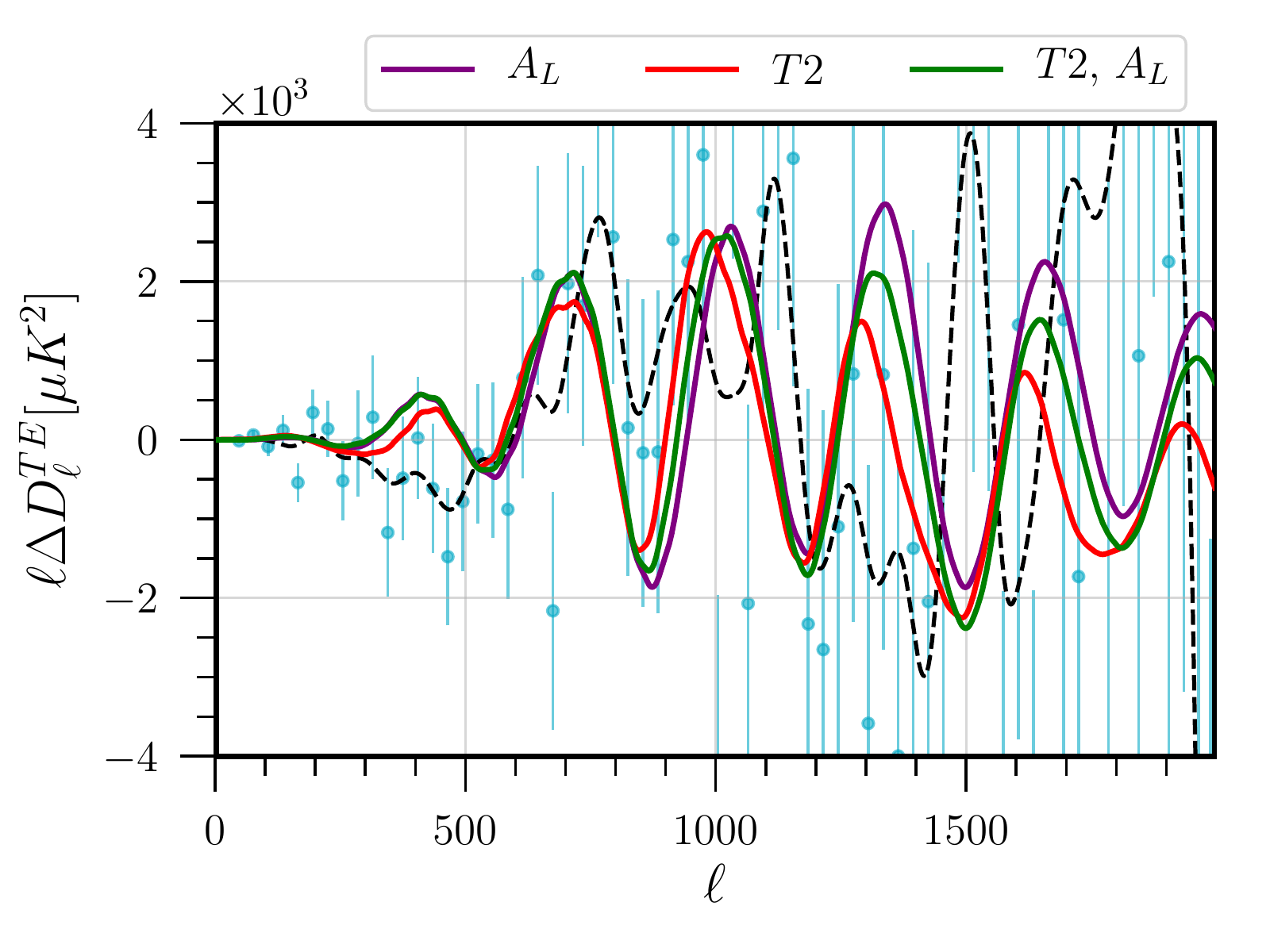}
\caption{Residuals of the CMB TT, EE and TE lensed power spectrum in the first, second and third row respectively. In light blue we show the residuals with the error bars of the Planck best fit against the data. The dotted black line represents the gaussian smoothed residuals from the data to provide an intuitive picture of the trend. In solid lines we present the residuals of the best-fit values for T1 and T2 (see table \ref{Tab:bestfitT1T2}) after subtracting the best fit value of $\Lambda$CDM provided by Planck. On the left, we see the residuals for the best fit of $A_L$ (purple), T1 (red) and $A_L$+T1 (green).  On the right, we see the residuals for the best fit of $A_L$ (purple), T2 (red) and $A_L$+T2 (green). Although both models give similar best fits to the eye, T1 which has a $k$ dependence frequency has a lower $\chi^2$ and hints to a signal generated in alternative scenarios to inflation, although with only marginal statistical significance. \label{fig:T1T2TE}}
\end{figure}

For illustrative purposes, we also plotted the residuals for best fit values of T1 and T2 (in table \ref{Tab:bestfitT1T2}) respectively on the left and right of Fig.~\ref{fig:T1T2TE}. In these plots we included the residuals of the $\Lambda$CDM best fit by Planck with error bars (light blue) and the gaussian smoothed residuals (dotted black) and the $\Lambda$CDM + $A_L$ (in purple). The residuals of T1 and T2 (in red lines) are similar to those of $\Lambda$CDM + $A_L$ for the TT power spectrum in the range $1000\lesssim\ell\lesssim 1500$. However, the possible degeneracy with the lensing smoothing is clearly broken by the EE and TE power spectra. Curiously, the best fit value for $A_L$ of T1 is $A_L=1.126$ while it is $A_L=1.184$ for T2. This result seems to suggest that the template T1 is more degenerate with the lensing anomaly. This may be the reason why it yields a better $\chi^2$ fit to the data when considering $\Lambda$CDM+T1 compared to $\Lambda$CDM+T2. We also recover Planck 2018 results \cite{Aghanim:2018eyx} regarding an oscillation with constant amplitude and frequency to explain the $A_L$ anomaly, that is template T1 with $\gamma=1$. In the appendix in table~\ref{Tab:allTT}, we see that T1 with fixed $\gamma=1$ does not reduce the lensing anomaly with $A_L=1.18_{-0.07}^{+0.07}$ at $68\%$ confidence level. Thus, we showed the importance in reducing the tension of a scale dependent frequency and/or amplitude of the oscillatory modulation. However, since neither T1 nor T2 do not significantly ease the tension, we also analyze both models with fixed $A_L=1$ in next section.

\begin{table}
\begin{center}
\begin{tabular}{|| c | c | c | c | c | c | c ||}
    \hline
    Model (TTTEEE  best fit) &  $A_L$  &   $A$   &   $\gamma$   &  $\omega$  & $\phi$  &  $\chi^2$  
    \\ [0.5ex]
    \hline\hline
    $\Lambda CDM$ + $A_L$ & $1.186$ &  - & - & - & - & $2760$
     \\[0.5ex]
    $\Lambda CDM$ + $A_L$ + T1  & $1.126$ & $0.009403$ & $1.184$ & $34.78$ & $-1.526$ & $2752$
    \\[0.5ex]
      $\Lambda CDM$ + T1  & - & $0.01624$ & $1.65$ & $36.79$ & $-0.4039$ & $2754$    \\[0.5ex]
    \hline
\end{tabular}
\end{center}
\begin{center}
\begin{tabular}{|| c | c | c | c | c | c ||}
   \hline
    Model (TTTEEE best fit) &   $A_L$   &   $B$   &  $\eta_f$  & $b_s$  &  $\chi^2$  
    \\ [0.5ex]
    \hline\hline
  $\Lambda CDM$ + $A_L$ + T2  & $1.184$ & $-0.03578$  & $121.1$ & $57.36$ & $2756$ \\ [0.5ex]

  $\Lambda CDM$ + T2  & - & $-0.07742$ & $122.3$ & $66.02$ & $2759$ \\ [0.5ex]
    \hline
\end{tabular}
\end{center}
\caption{Best fit values for the modulation T1 (top) and T2 (bottom) using the full TTTEEE+low E Planck likelihood. On the top table, we see that the best fit value for $A_L$ has been reduced. However, on the bottom table we find that $A_L$ has been not reduced although the $2\sigma$ constraints in table~\ref{Tab:meanT1T2} prefer a lower value. The best fit for the TT+low E likelihood is presented in Tab.~\ref{Tab:allTT} in App.~\ref{app:T1extra}. \label{Tab:bestfitT1T2}}
\end{table}

\paragraph{Implications for alternative scenarios to inflation:} Let us discuss the implications of our results as possible signatures of alternatives scenarios to inflation. From table \ref{Tab:meanT1T2}, we see that T1 (with 4 extra free parameters) provides the lowest $\chi^2$ fit to the data compared to T2 (3 extra free parameters) and $A_L$ (1 extra free parameter). With respect to $\Lambda$CDM we find that $\Delta\chi^2_{T1}\equiv\chi^2_{T1}-\chi^2_{\Lambda{\rm CDM}}=-13$, $\Delta\chi^2_{T2}=-8$ and $\Delta\chi^2_{A_L}=-7$. These results seem to suggest that the oscillatory modulation T1 provides a more plausible explanation for the $A_L$ anomaly than T2, although the latter has one fewer free parameter.\footnote{We could have use the more general template of Ref.~\cite{Achucarro:2014msa} introducing a phase to template T2 \eqref{eq:T2}. However, this would double the number of free parameters and we preferred to keep the number of free parameters to the minimum.}  In this regard, we use the Akaike Information Criterion\footnote{The AIC is defined by \cite{akaike}
\begin{align}
{\rm AIC}\equiv2n-2\ln{\cal L}\,,
\end{align}
where $n$ is the number of parameters and ${\cal L}$ is the maximum likelihood. We also used that $-2\ln{\cal L}\sim\chi^2$ \cite{Liddle:2009xe}. The model with the lowest AIC constitutes a ``better '' candidate to explain the data.} (AIC) \cite{akaike,Liddle:2004nh,Liddle:2007fy} as an analytical example of a statistical model selection criterion \cite{Liddle:2009xe}. The results of the AIC should be viewed as a rough comparison between models. We obtain that the difference with respect to $\Lambda$CDM using the TTTEEE+low E likelihood is given by
\begin{align}\label{eq:AIC}
\Delta {\rm AIC}=\left\{
\begin{aligned}
&-5 \qquad (\Lambda{\rm CDM}+A_L)\\
&-5	\qquad (\Lambda{\rm CDM}+{\rm T1})\\
&-2 \qquad (\Lambda{\rm CDM}+{\rm T2})
\end{aligned}
\right.\,,
\end{align}
where we defined $\Delta{\rm AIC}[i]={\rm AIC}[i]-{\rm AIC}[{\Lambda{\rm CDM}}]$ and $i$ stands for the model being compared. The values of the AIC suggest that model T1 provides an equally reasonable fit to the data compared to $A_L$. Interestingly, if we include the lensing likelihood we find that the AIC value for $A_L$ is reduced to $\Delta{\rm AIC}[A_L]=-3$ while the AIC value for T1 remains the same at $\Delta{\rm AIC}[{\rm T1}]=-5$ (see table~\ref{Tab:meanT1lensing}). It should be noted that according to Jeffrey's scale \cite{Liddle:2007fy} a model with $|\Delta {\rm AIC}|>5$, should be view as a ``strong'' candidate and $|\Delta {\rm AIC}|>10$ as a ``decisive'' candidate compared to the model with a higher AIC. Thus, model T1 constitute an attractive candidate to be tested by future observations \cite{Huang:2012mr,Meerburg:2015owa,Chen:2016vvw,Ballardini:2016hpi,Chen:2016zuu, Xu:2016kwz,Ballardini:2017qwq,Beutler:2019ojk}.

Now, let us remind the reader that the $k^\gamma$ dependence in the frequency of oscillatory modulation T1 may be directly related to the power-law index $\alpha$ of the scale factor in a power-law universe \eqref{eq:powerlaw}. This result of Sec.~\ref{sec:imprint} was based on the assumption that the mass of the heavy field is constant. Furthermore, we argued that a constant mass might be favored by Occam's razor (or quantitatively by AIC) due to the reduced number of parameters compared to a time-dependent mass. Interestingly, we obtain that the value of power-law index of the frequency of T1 at $68\%$ confidence level is (see tables \ref{Tab:meanT1T2} and \ref{Tab:allTT})
\begin{align}\label{eq:gamma}
\gamma=\left\{
\begin{aligned}
&1.69_{-0.25}^{+0.25} \qquad ({\rm TTTEEE+low \,E})\,,\\
&2.10_{-0.49}^{+0.51} \qquad ({\rm TT+low \,E})\,.
\end{aligned}
\right.
\end{align}
The inclusion of lensing does not change the constraints on $\gamma$. We also found that in the TTTEEEE case $\gamma\neq 1$ is excluded close to $3\sigma$. Considering that T1 has a low AIC value, our result Eq.~\eqref{eq:gamma} shows that the data seems to prefer a $k$-dependent frequency of the oscillatory modulation to the primordial power spectrum. This implies that, with the constant mass assumption, the oscillatory feature was generated during a contracting phase where the scale factor evolved as
\begin{align}
a\propto \tau^\alpha \quad {\rm with} \quad \alpha=1.45\pm0.53\,.
\end{align}
In other words, the results favor a primordial universe that was dominated by a fluid with an equation of state between the dust and radiation, during the epoch when the density perturbations were generated. Namely,
\begin{align}
w=0.13\pm0.17\,.
\end{align}
At present, we are not aware of a particular bouncing model with $w\approx 0.13$. Nevertheless, our result is compatible within $2\sigma$ with alternative scenarios to inflation such as matter bounce
\cite{Wands:1998yp} and radiation bounce \cite{Boyle:2018tzc} where, respectively, $\gamma=3/2$ ($\alpha=2$, $w=0$) and $\gamma=2$ ($\alpha=1$, $w=1/3$). Ekpyrotic models \cite{Khoury:2001wf} of the very early universe does not provide an explanation for such a signal as they require $\gamma\gg 1$ ($0<\alpha\ll1$, $w\gg 1$), i.e. a very slow contraction scenario.

\begin{table}
\begin{center}
\begin{tabular}{|| c | c | c | c | c | c | c ||}
    \hline
    \begin{tabular}[c]{@{}c@{}} Model (TTTEEE \\+ lensing 68\% confidence)\end{tabular}  &   $A_L$   &   $A$   &   $\gamma$   &  $\omega$  & $\phi$  &  $\chi^2$  
    \\ [0.5ex]
    \hline\hline
    $\Lambda CDM$  & - & - & - & - & - & $2777$
     \\[0.5ex]
    $\Lambda CDM$ + $A_L$ & $1.12_{-0.06}^{+0.05}$ &  - & - & - & - & $2772$
     \\[0.5ex]
    $\Lambda CDM$ + $A_L$ + T1 & $1.10_{-0.06}^{+0.05}$ & $0.012_{-0.005}^{+0.005}$ & $1.53_{-0.41}^{+0.40}$ & $37_{-3}^{+4}$ & $-0.6_{-0.5}^{+0.6}$ & $2763$
    \\[0.5ex]
      $\Lambda CDM$ + T1 & - & $0.015_{-0.005}^{+0.005}$ & $1.69_{-0.29}^{+0.25}$ & $37_{-3}^{+3}$ & $-0.5_{-0.4}^{+0.5}$ & $2764$
    \\[0.5ex]
  \hline
\end{tabular}
\end{center}
    \caption{Constraints on  T1 using the full TTTEEE + low E + lensing CMB likelihood from Planck 2018 \cite{Aghanim:2018eyx}. For comparison we included the constraints on the $A_L$ parameter with and without T1. Note that that including the lensing likelihood reduces the $A_L$ tension in $\Lambda$CDM + $A_L$ yielding $A_L=1$ consistent within $2\sigma$. Adding the oscillatory feature T1 further reduces the anomaly. Also we find that for $\Lambda$CDM + T1 the value $\gamma=1$ is excluded close to $3\sigma$. Note how introducing lensing does not significantly change the main results of table \ref{Tab:meanT1T2} using the TTTEEE+low E likelihood. This time we find that $\Lambda$CDM + T1 is more favored by the AIC having $\Delta {\rm AIC}[{\rm T1}]=-5$ than $\Lambda$CDM + $A_L$ with $\Delta{\rm AIC}[{A_L}]=-3$. \label{Tab:meanT1lensing}}
\end{table}

In this paper, we only analyzed the parameter space for alternative-to-inflation scenarios. It would be interesting to briefly review and compare the results of the related feature models in the inflation scenario. The residuals of the power spectrum at $\ell>700$  measured by Planck can also be fit by several inflationary feature models, see e.g.~\cite{Pahud:2008ae,Meerburg:2013cla,Meerburg:2013dla,Hu:2014hra,Zeng:2018ufm,Akrami:2018odb}. In particular, Ref.~\cite{Akrami:2018odb} showed that the inflationary resonance models \cite{Chen:2008wn,Silverstein:2008sg,Flauger:2009ab}, where the oscillatory signals are generated by periodic features in the inflationary potential,  can reduce the $\chi^2$  by $-11$ relative to LCDM with 3 extra parameters. This gives the same AIC as the candidate analyzed in this paper, although these features are not due to the PSCs. On the other hand, the same Planck residuals may also be fit by other models where the signals are generated by the PSCs in the inflation scenario \cite{Chen:2014joa,Chen:2014cwa}.  Detailed comparison with the Planck 2018 data is in progress \cite{ChenHazra}.

As we can see, both the inflation and non-inflation candidates mentioned above so far all only have marginal statistical significance, and the Planck power spectrum residuals are still consistent with statistical fluctuations. However, with future CMB polarization experiments and large scale structure surveys \cite{Huang:2012mr,Meerburg:2015owa,Chen:2016vvw,Ballardini:2016hpi,Chen:2016zuu,Xu:2016kwz, Ballardini:2017qwq,Beutler:2019ojk,Ballardini:2019tuc}, these candidates can be further tested, and can be falsified or distinguished from each other.

\section{Conclusions\label{sec:conclusions}}
Future galaxy surveys, together with future CMB experiments, will be able to place new constraints on oscillatory features in the primordial power spectrum \cite{Huang:2012mr,Meerburg:2015owa,Chen:2016vvw,Ballardini:2016hpi,Chen:2016zuu, Xu:2016kwz,Ballardini:2017qwq,Beutler:2019ojk,Ballardini:2019tuc}. It has also been found that such oscillatory features can be modelled to non-linear evolution of matter and could be seen in the BAO \cite{Vlah:2015zda,Beutler:2019ojk,Vasudevan:2019ewf,Ballardini:2019tuc}. Interestingly, the latest analysis of CMB anisotropies by Planck 2018 \cite{Aghanim:2018eyx} found that there is $10\%$ more lensing than expected in the CMB power spectrum, with $A_L\neq1$ at $2.8\sigma$. A plausible explanation for such lensing anomaly is an oscillatory modulation in the primordial power spectrum since they are degenerate with the smoothing effect of lensing \cite{Hazra:2017joc}, in particular if they have a frequency similar to the acoustic peaks. It should be noted that polarization data by SPTpol shows a preference for lower values of the lensing anomaly, although within $1.4\sigma$ with $\Lambda$CDM expectations \cite{Aghanim:2018eyx,Simard:2017xtw,Motloch:2019gux}. Nevertheless, the lensing anomaly may hint to oscillatory features in the primordial power spectrum regardless of whether they significantly ease the tension or not.

In this paper, we analyzed oscillatory modulations caused by two types of feature models: the sharp feature signals which have constant frequency for both inflation and alternative-to-inflation models; and the PSC signals generated by quantum-mechanically oscillating massive fields in alternative scenarios to inflation, where the frequency acquires a power-law dependence in $k$ \cite{Chen:2011zf,Chen:2018cgg}. We contrasted the candidate (T2) \eqref{eq:T2} from Ref.~\cite{Domenech:2019cyh} to explain the $A_L$ anomaly with the latest CMB data by Planck \cite{Aghanim:2018eyx}. We also used the modulations (T1) \eqref{eq:T1} of Ref.~\cite{Chen:2018cgg} generated in alternative scenarios to inflation, which includes the matter or radiation bounce cosmology \cite{Wands:1998yp,Boyle:2018tzc}. In the former, the frequency of the oscillations is constant, i.e. it goes as $\Delta P_{\cal R}\sim\sin(\omega k)$ where $k$ is the wavenumber, and it is generated by a sharp feature during inflation. In the latter, the frequency goes as a power-law of $k$, namely $\Delta P_{\cal R}\sim\sin(\omega k^\gamma)$, and its originates by a massive scalar field in a contracting universe, e.g. $\gamma=3/2$ for matter bounce and $\gamma=2$ for radiation bounce.

We used Boltzmann system solver \texttt{CLASS} \cite{Blas:2011rf} to compute the resulting CMB anisotropies and did the Bayesian exploration with the MonteCarlo sampler \texttt{MontePython} \cite{Audren:2012wb,Brinckmann:2018cvx} varying the six standard cosmological and nuisance/foreground parameters.  Furthermore, we use the full TTTEEEE+low E CMB likelihood, also including low T, from Planck 2018 \cite{Aghanim:2018eyx,Aghanim:2019ame} and all results will refer to TTTEEE unless stated otherwise. At first, we did not consider the lensing likelihood due to the fact that the anomaly is mainly preferred by the temperature and polarization power spectra. Thus, we studied models that would reduce the anomaly using the power spectra alone. Also, we restricted our analysis to the first relevant peak in the likelihood near the frequency of the acoustic peaks due to the appearance of multimodality, i.e. nearby peaks in the likelihood. Later we considered the effects of including the lensing likelihood for the case with lowest $\chi^2$. The main results can be found in Sec.~\ref{sec:results}, tables \ref{Tab:meanT1T2} and \ref{Tab:meanT1lensing} and Figs.~\ref{fig:ALgamma} and \ref{fig:1d2dT1T2}. First, using the TTTEEE+low E likelihood we found that the oscillatory modulations T1 \eqref{eq:T1} (4 extra free parameters) and T2 \eqref{eq:T2} (3 extra free parameters) reduce the lensing anomaly to a statistical significance slightly below $2\sigma$, in contrast with Planck 2018 results (1 extra free parameter) where $A_L\neq 1$ at $2.8\sigma$. In particular we obtained that in the model $\Lambda{\rm CDM}\,+\,A_L\,+\,{\rm T1}$ at $68\%$ confidence
\begin{align}
A_L=\left\{
\begin{aligned}
&1.14^{+0.07}_{-0.07}\qquad ({\rm TTTEEE+ low E})\,,\\
\end{aligned}
\right.
\end{align}
Thus, compared to the $2.8\sigma$ in $\Lambda CDM$ the model $\Lambda{\rm CDM}\,+\,A_L\,+\,{\rm T1}$ only mildly reduces the tension to $2\sigma$. For this reason, we also analyze the model T1 with fixed $A_L=1$. We also recovered that an oscillation with constant amplitude and frequency does not reduce the tension \cite{Aghanim:2018eyx} (see table \ref{Tab:allTT}). Thus, our work shows the importance of the $k$-dependence in the amplitude and frequency to ease the lensing anomaly. Second, the amplitude of the oscillations is non-zero close to $3\sigma$ for T1 and $2\sigma$ for T2. When considering the models T1 and T2 with fix $A_L=1$, we obtained that it presents the lowest value of $\chi^2$ with $\Delta\chi^2_{T1}=-13$ with respect to $\Lambda$CDM while T2 has $\Delta\chi^2_{T2}=-8$.
We also used the AIC as a statistical criterion for model selection. We obtained that $\Delta{\rm AIC}=-5$ for both $A_L$ and T1 compared to $\Lambda$CDM, while T2 is not preferred with $\Delta{\rm AIC}=-2$. The inclusion of the lensing likelihood lowers the AIC value for $A_L$ to $\Delta{\rm AIC}[A_L]=-3$ while it remains the same for T1 with $\Delta{\rm AIC}[{\rm T1}]=-5$. Thus, although only mildly related to the lensing anomaly, the model T1 constitutes a falsifiable new candidate to be tested by future data.

In Sec.~\ref{sec:imprint} we have argued that the modulation T1 due to a massive scalar field may be a probe of the evolution of the primordial universe. If the mass of the scalar field is constant, the $k^\gamma$ dependence of the frequency of the oscillatory modulation in the primordial spectrum is directly related to the power-law index $\alpha$ of the scale factor. In other words, the value of the parameter $\gamma$ distinguishes different scenarios of the primordial universe. In our analysis with TTTEEE+low E and TTTEEE+low E+lensing we found that $\gamma=1.69_{-0.25}^{+0.25}$ at $68\%$ confidence level, and it is above $\gamma=1$, typical of a sharp feature, close to $3\sigma$. Under the constant mass assumption and considering that such model also presents a low AIC value, our results suggest that during the generation of the feature the universe was contracting and dominated by fluid with an equation of state between dust ($w=0$) and radiation ($w=1/3$), explicitly at $68\%$ confidence
\begin{align}
w=0.13\pm0.17 \qquad ({\rm TTTEEE+ low E})\,,
\end{align}
where the constraint is unchanged if we include the lensing likelihood. This value of the equation of state is compatible within $2\sigma$ with the matter bounce and radiation bounce cosmology. In contrast, this candidate is incompatible with oscillatory modulations of the same nature in the ekpyrotic scenario, as it requires $w\gg1$. On the other hand, the AIC value of this non-inflationary candidate is similar to some feature model candidates in the inflation scenario. In some sense, this might be the first test of the evolutionary history of the scale factor of the primordial universe and a comparison between inflation and its alternatives, as an independent and complimentary approach to the method of primordial gravitational waves \cite{Brandenberger:2011eq,Ade:2015tva,Kamionkowski:2015yta,Ade:2018iql}.

\section*{Acknowledgments}

We thank R.~Brandenberger and T.~Stark for useful feedback. G.D. would like to thank E.~Bittner for his constant IT help, J.~Fedrow for helpful discussions, C.~Fidler for kindly providing the \texttt{CLASS} code with the modification for the $A_L$ parameter, A.~G\'{o}mez-Valent for his help with \texttt{MontePython} and for useful comments on the manuscript and J.~Torrado for the discussions on multimodality. G.D. thanks the UC Berkeley cosmology group for their hospitality while this work was being done. G.D. and X.C. would also like to thank D.~Hazra and Y.~Wang for useful correspondence. G.D. was partially supported by the DFG Collaborative Research center SFB 1225 (ISOQUANT) and the European Union’s Horizon 2020 research and innovation programme (InvisiblesPlus) under the Marie Sk{\l}odowska-Curie grant agreement No 690575. A.L. acknowledges support by the Black Hole Initiative at Harvard University which is funded by JTF and GBMF grants. M.K. was supported in part by NASA Grant No. NNX17AK38G, NSF Grant No. 1818899, and the Simons Foundation.

\appendix

\section{Additional results for the modulation T1 \label{app:T1extra}}

Using the full TTTEEE+low E Planck likelihoods on template T1 for fixed values of the power-law index $\gamma$ we find table \ref{Tab:meanT1T2all}.

\begin{table}[H]
\begin{center}
\begin{tabular}{|| c | c | c | c | c | c | c | c | c | c | c | c | c | c ||}
    \hline
    Model (best fit) & $10^{+9}A_s$ & $n_s$ & $A$ & $\omega$ & $\gamma$ & $\phi$ & $100~\Omega_{b }h^2$
    \\ [0.5ex]
    \hline\hline
    $\Lambda CDM$ & $2.076$ & $0.9636$ & - & - & - & - & $2.233$ 
     \\[0.5ex]
    $\Lambda CDM$ + $T1$ & $2.106$ &  $0.9608$  & $0.01624$ & $36.79$ & $1.65$ & $-0.4039$ & $2.224$ 
     \\[0.5ex]
    \hline
\end{tabular}
\end{center}
\begin{center}
\begin{tabular}{|| c | c | c | c | c | c | c | c | c | c | c | c | c | c ||}
    \hline
    Model (best fit) & $\Omega_{cdm }h^2$ & $\Omega_{\Lambda }$ & $H_0$ & $\sigma_8$ & $100~\theta_{s }$ & $z_{reio}$
    \\ [0.5ex]
    \hline\hline
    $\Lambda CDM$ & $0.1193$ & $0.6888$ & $67.63$ & $0.8037$ & $1.042$ & $7.489$
     \\[0.5ex]
    $\Lambda CDM$ + $T1$ & $0.1226$ & $0.6703$ & $66.43$  & $0.8199$ & $1.042$ & $7.697$
     \\[0.5ex]
    \hline
\end{tabular}
\end{center}
    \caption{Best fit values for the modulation T1 \eqref{eq:T1} using the the TTTEEE+low E Planck likelihood including some cosmological parameters. We also wrote the results for $\Lambda$CDM for easier comparison. See how the paremeters $\Omega_{cdm}h^2$ and $z_{reio}$ are change by almost $3\%$ with respect to the base $\Lambda$CDM. \label{Tab:bestfitall}}
\end{table}

\begin{table}[H]
\begin{center}
\begin{tabular}{|| c | c | c | c | c | c | c ||}
    \hline
   \begin{tabular}[c]{@{}c@{}}  Model (TTTEEE \\68\% confidence)\end{tabular} &   $A_L$   &   $A$   &   $\gamma$   &  $\omega$  & $\phi$  &  $\chi^2$  
    \\ [0.5ex]
    \hline\hline
    $\Lambda CDM$  & - & - & - & - & - & $2767$
     \\[0.5ex]
    $\Lambda CDM$+$A_L$ & $1.18_{-0.07}^{+0.07}$ &  - & - & - & - & $2760$
     \\[0.5ex]
    $\Lambda CDM$+$A_L$+T1 ($\alpha$ free) & $1.14_{-0.07}^{+0.07}$ & $0.012_{-0.005}^{+0.005}$ & $1.56_{-0.38}^{+0.18}$ & $36_{-3}^{+3}$ & $-0.8_{-0.5}^{+0.6}$ & $2752$
    \\[0.5ex]
      $\Lambda CDM$+T1 ($\alpha$ free) & - & $0.015_{-0.005}^{+0.005}$ & $1.69_{-0.25}^{+0.25}$ & $37_{-3}^{+2}$ & $-0.5_{-0.3}^{+0.5}$ & $2754$
    \\[0.5ex]
   $\Lambda CDM$+$A_L$+T1 ($\alpha\to\infty$) & $1.18_{-0.07}^{+0.07}$ & $0.011_{-0.004}^{+0.004}$ & $1$ & $33_{-3}^{+3}$ & $-0.04_{-0.4}^{+0.3}$ & $2753$
  \\[0.5ex]
   $\Lambda CDM$+T1 ($\alpha\to\infty$) & -  & $0.012_{-0.004}^{+0.005}$ & $1$ & $36_{-2}^{+3}$ & $-0.3_{-0.4}^{+0.3}$ & $2756$
  \\[0.5ex]
  $\Lambda CDM$+$A_L$+T1 ($\alpha=2$) & $1.15_{-0.08}^{+0.08}$ & $0.012_{-0.005}^{+0.005}$ & $3/2$ & $37_{-3}^{+3}$ & $1.2_{-0.2}^{+0.2}$ & $2749$ \\[0.5ex]

  $\Lambda CDM$+T1 ($\alpha=2$) & - & $0.015_{-0.005}^{+0.005}$ & $3/2$  & $36_{-2}^{+2}$ & $1.2_{-0.2}^{+0.2}$ & $2754$ \\[0.5ex]

  $\Lambda CDM$+$A_L$+T1 ($\alpha=1$) & $1.13_{-0.07}^{+0.07}$ & $0.011_{-0.006}^{+0.006}$ & $2$ & $40_{-3}^{+4}$ & $-0.01_{-0.31}^{+0.17}$ & $2755$ \\[0.5ex]

  $\Lambda CDM$+T1 ($\alpha=1$) & - & $0.016_{-0.005}^{+0.005}$ & $2$  & $40_{-2}^{+2}$ & $-0.06_{-0.19}^{+0.14}$ & $2754$ \\[0.5ex]
    \hline
\end{tabular}
\end{center}
    \caption{Constraints on T1 \eqref{eq:T1} fixing the power-law index $\gamma$ to see the effect on $A_L$ using the full TTTEEE + low E CMB likelihood from Planck 2018 \cite{Aghanim:2018eyx}. For comparison we included the constraints on the $A_L$ parameter with and without T1. See how a value of $\gamma=2$ is the one that reduces the $A_L$ parameter the most. This is also compatible with the fact that $TT$ prefers a value of $\gamma\sim 2$ (see table~\ref{Tab:allTT} and Fig.~\ref{fig:1d2dT1TT}).\label{Tab:meanT1T2all}}
\end{table}

\begin{figure}[H]
\includegraphics[width=0.49\columnwidth,valign=m]{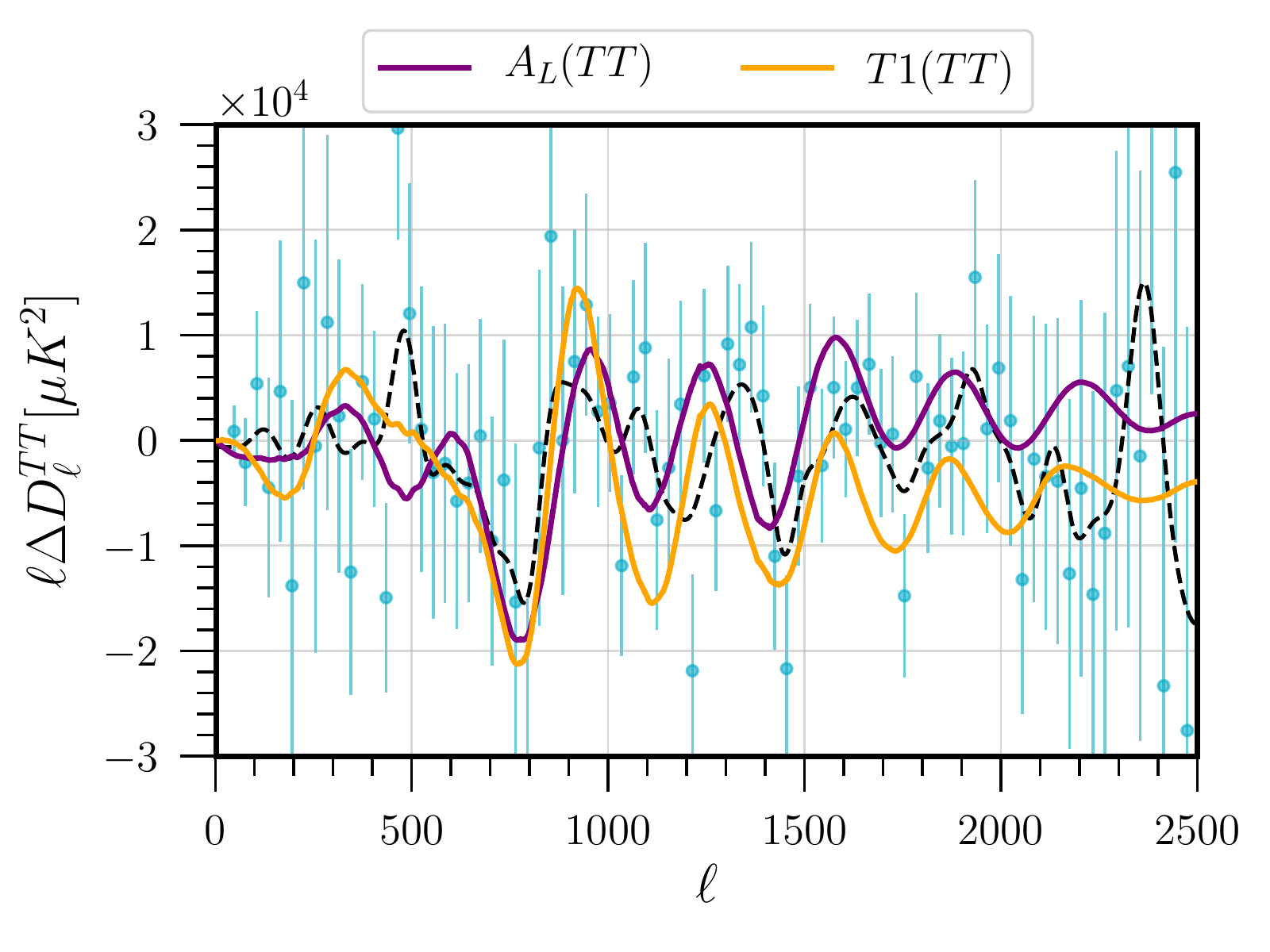}
\caption{ Residuals of the CMB TT lensed power spectrum for the model T1. In light blue we show the residuals with the error bars of the Planck best fit against the data. The dotted black line represents the gaussian smoothed residuals from the data to provide an intuitive picture of the trend. In orange lines we see the residuals of the best-fit values for T1 using TT+low E of table \ref{Tab:allTT} after subtracting the best fit value of $\Lambda$CDM provided by Planck. \label{fig:T1TT}}
\end{figure}

\begin{figure}[H]
\includegraphics[width=0.60\columnwidth,valign=m]{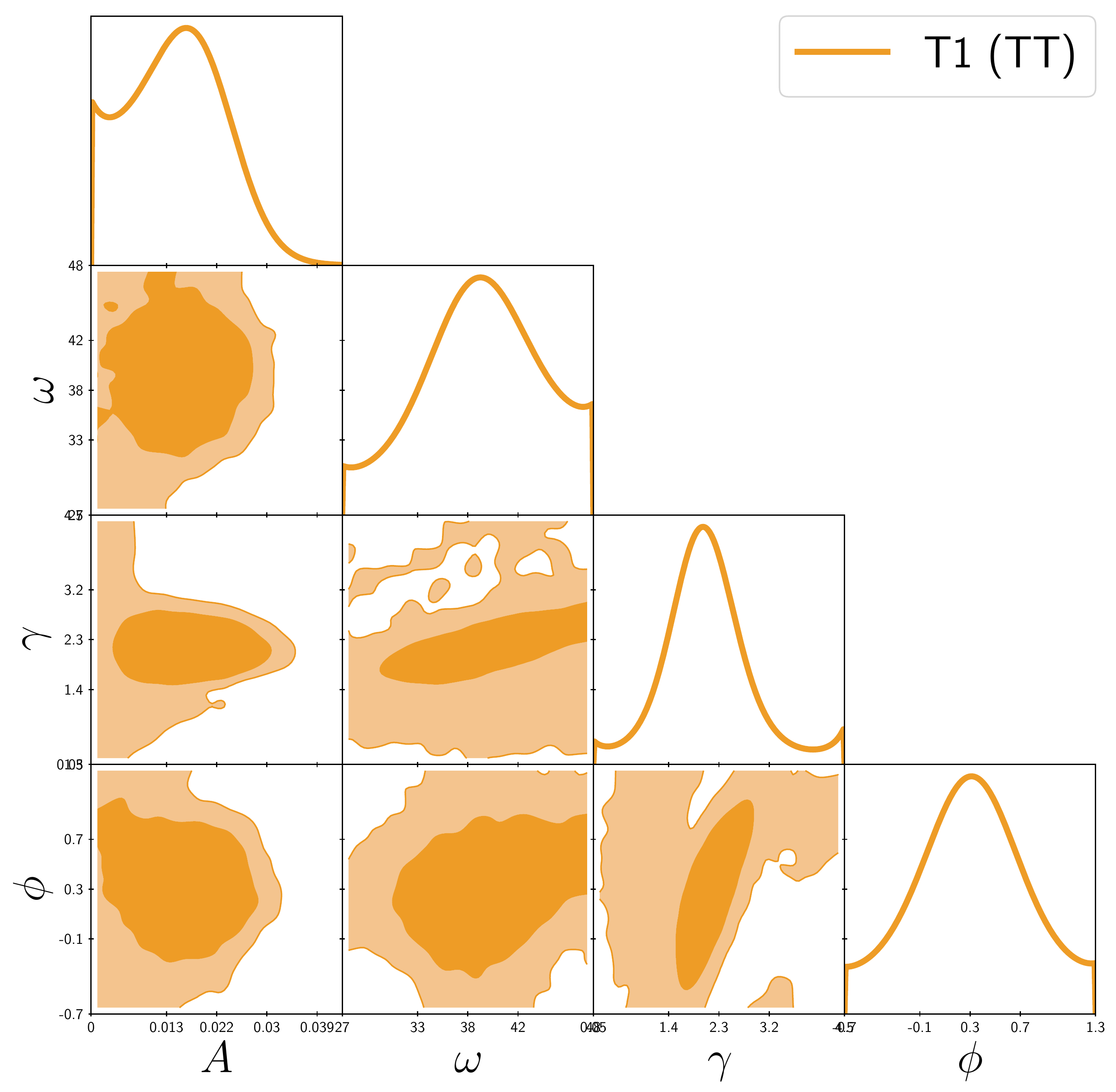}
\caption{ 1D and 2D marginalized likelihoods with the $1\sigma$ and $2\sigma$ contours for the modulation T1 \eqref{eq:T1} using the TT+low E Planck likelihood. We restricted our study to the first relevant peak near the frequency of the acoustic peaks due to multimodality. Using only TT+low E the power-law index $\gamma$ prefers a value close to $2$ which corresponds to radiation bounce cosmology ($\alpha=1$). It should be noted that the seemingly growth of the likelihood near the prior boundaries in $\gamma$ are numerical effects as we checked by changing the priors in $\gamma$. A detailed study of multimodality in the parameter space is left for future work. \label{fig:1d2dT1TT}}
\end{figure}

Now, if we only use the TT+low E Planck likelihoods with template T1 we obtain table \ref{Tab:allTT}. The 1D and 2D marginalized likelihoods are shown in Fig.~\ref{fig:1d2dT1TT} and an illustrative plot of the best fit values in Fig.~\ref{fig:T1TT}.

\begin{table}[H]
\begin{center}
\begin{tabular}{|| c | c | c | c | c | c | c ||}
    \hline
    Model (TT 68\% confidence)  &   $A_L$   &   $A$   &   $\gamma$   &  $\omega$  & $\phi$  &  $\chi^2$  
    \\ [0.5ex]
    \hline\hline
    $\Lambda CDM$  & - & - & - & - & - & $1179$
     \\[0.5ex]
    $\Lambda CDM$ + $A_L$ & $1.25_{-0.10}^{+0.10}$ &  - & - & - & - & $1173$
     \\[0.5ex]
      $\Lambda CDM$ + T1 & - & $0.015_{-0.009}^{+0.009}$ & $2.10_{-0.49}^{+0.51}$ & $39_{-5}^{+5}$ & $0.3_{-0.4}^{+0.5}$ & $1170$\\[0.5ex]
  \hline
\end{tabular}
\end{center}
\begin{center}
\begin{tabular}{|| c | c | c | c | c | c | c ||}
    \hline
    Model (TT best fit)  &   $A_L$   &   $A$   &   $\gamma$   &  $\omega$  & $\phi$  &  $\chi^2$  
    \\ [0.5ex]
    \hline\hline
    $\Lambda CDM$ + $A_L$ & $1.247$ &  - & - & - & - & $1173$
     \\[0.5ex]
      $\Lambda CDM$ + T1 & - & $0.01832$ & $1.846$ & $36.17$ & $-0.03583$ & $1170$\\[0.5ex]
  \hline
\end{tabular}
\end{center}
    \caption{Top: Constraints on T1 \eqref{eq:T1} using the only TT + low E CMB likelihood from Planck 2018 \cite{Aghanim:2018eyx}. For comparison we included the constraints on the $A_L$ parameter with and without T1. See how $TT$ prefers a value of $\gamma\sim 2$. Bottom: Best fit values for the modulation T1 \eqref{eq:T1} using the only the TT+low E Planck likelihood. See how the best fit value for $A_L$ has been reduced.\label{Tab:allTT}}
\end{table}

\bibliographystyle{jhep}
\bibliography{biblio.bib}

\end{document}